\documentclass[preprint]{aastex}

\usepackage{natbib}
\usepackage{graphicx}
\usepackage[loose]{units}

\newcommand{\beq}{\begin{equation}}
\newcommand{\eeq}{\end{equation}}
\newcommand{\un}{\unit}

\begin{document}

\title{Wavelets with ridges: a high-resolution representation of cataclysmic variable time-series}
\author{Claire Blackman}
\affil{Department of Economics, Royal Holloway, University of
London} \email{claire.blackman@rhul.ac.uk}

\begin{abstract}
Quasi-periodic oscillations and dwarf nova oscillations occur in
dwarf novae and nova-like variables during outburst and occasionally
during quiescence, and have analogues in high-mass X-ray binaries
and black-hole candidates. The frequent low coherence of
quasi-period oscillations and dwarf nova oscillations can make
detection with standard time-series tools such as periodograms
problematic. This paper develops tools to analyse quasi-periodic
brightness oscillations. We review the use of time-frequency
representations in the astronomical literature, and show that
representations such as the Choi-Williams Distribution and
Zhao-Atlas-Marks Representation, which are best suited to high
signal-to-noise data, cannot be assumed a priori to be the best
techniques for our data, which have a much higher noise level and
lower coherence. This leads us to a detailed analysis of the
time-frequency resolution and statistical properties of six
time-frequency representations. We conclude that the wavelet
scalogram, with the addition of wavelet ridges and maxima points, is the most effective
time-frequency representation for analysing quasi-periodicities in
low signal-to-noise data, as it has high time-frequency resolution,
and is a minimum variance estimator.

We use the wavelet ridges method to re-analyse archival data from VW Hyi, and find 62 new
QPOs and 7 new long-period DNOs. Relative to previous
analyses, our method substantially improves the detection rate for
QPOs.
\end{abstract}


\section{Introduction}

Cataclysmic variable (CV) light curves frequently contain multiple
(quasi-)periodic components covering a wide frequency range, from
\un[3]{s} dwarf nova oscillations (DNOs) \citep{kn.Wo02a} through
$\sim$\un[100]{s} longer period DNOs (lpDNOs) \citep{kn.Wa03} to
$\sim$\un[5000]{s} quasi-periodic oscillations (QPOs)
\citep{kn.Pa77}. Each of these components can vary in period,
amplitude and phase on short time-scales, against a high level of
background noise. DNOs, with $Q=|\frac{dP}{dt}|^{-1}\approx10^4 - 10^6$, are generally more coherent than QPOs ($Q\approx1 - 10$), but both are seldom visible directly in the light
curve, and the periodogram, which is designed for stationary
phenomena, can fail to detect them \citep[see for
example][]{kn.Ro79}. Thus in order to study these and similar
phenomena it is important to find a tool that detects them reliably.
An investigation of time-frequency representations, which allow a
non-stationary signal to be visualized in the time and frequency
domains simultaneously, and have been successfully used in other
areas of astronomy, would seem to be an intuitive and instructive
place to begin.

The most common time-frequency representation in use in astronomy is
the wavelet transform, either in the form of the weighted wavelet
Z-transform (WWZ) of \citet{kn.Fo96} or the wavelet scalogram. We
will focus on the wavelet scalogram as our data are regularly
spaced, and the WWZ is designed for irregularly spaced data and has
poorer time-frequency resolution. The wavelet scalogram has been
used to analyse, amongst other astronomical phenomena, SR-type stars
\citep{kn.Ch00,kn.Ki06}, CVs \citep{kn.Fr98,kn.Ha02} and planetary
nebulae \citep{kn.Go03}. The significance testing method of
\citet{kn.To98} for the wavelet scalogram is used mainly in solar
astrophysics, where it has been applied to magnetic field
measurements \citep{kn.Bo02}, coronal variability \citep{kn.Ma02},
high-frequency oscillations in coronal loops \citep{kn.Ka02c} and
chromospheric UV oscillations \citep{kn.Mc04}.

Several other time-frequency representations have also been used in
astronomy. The Gabor spectrogram has been applied to synthetic
multi-component signals and roAP stars \citep{kn.Bo95} and QPOs in
both the atoll source 4U 1820-30 \citep{kn.Be04} and Her X-1
\citep{kn.OB01}. The Choi-Williams distribution was applied to
Wolf-Rayet stars \citep{kn.Ma98a} and the Cone-Kernel Representation
to irregular pulsators \citep{kn.Bu04} and T UMi \citep{kn.Sz03}.

There are several key differences between our data and those
analysed in the literature which make the choice of the most
appropriate time-frequency representation for our analysis unclear.
Firstly, in all of the papers mentioned above (and, indeed, all
those in the literature) either visual inspection of the light curve
or calculation of the periodogram is the primary source of
periodicity detection; the time-frequency representation serves
merely as an adjunct. For our signals, which have lower
signal-to-noise ratios than those in the literature and which cannot
be seen directly in the light curve or the periodogram, the
time-frequency representation will be the primary means of
detection, and hence it is imperative that we be able to test the
significance of any detections statistically. Apart from
\citet{kn.Bo95}, the only papers conducting significance testing of
time-frequency representations are those using the \citet{kn.To98}
wavelet scalogram method.

Our analysis also requires a broader frequency range than
any discussed in the literature, as well as very high frequency
resolution, particularly in the QPO frequency range as QPOs often
appear as closely spaced harmonics. In addition, since both DNOs
and QPOs are intermittent phenomena that can change their
characteristics on a time-scale of \un[10]{s} of seconds, we need high time resolution over the entire frequency range. Thus
time-frequency representations with higher time-frequency resolution
would be preferable for our analysis. Comparisons of the wavelet
scalogram, Gabor spectrogram, Choi-Williams distribution and the
Cone-Kernel Representation show that, at least for relatively large
amplitude variables stars such as irregular pulsators or Mira-types,
the Choi-Williams distribution and Cone-Kernel Representation can
give better time-frequency resolution than the wavelet scalogram
\citep[e.g.][]{kn.Ko97,kn.Bu01,kn.Ki02a}.

A plethora of papers investigate the effectiveness of different
time-frequency representations using various synthetic signals (see
for example \citet{kn.Go91}, \citet{kn.Ca92},
\citet{kn.Sz94}, \citet{kn.Bo95}, \citet{kn.Li99}, \citet{kn.Hl95},
\citet{kn.Bu01}, \citet{kn.Ki02a}, \citet{kn.Th03}, \citet{kn.Sz03}
or \citet{kn.Fi04}), but as none of these synthetic data have features comparable to those of our data we cannot necessarily
use their conclusions.

Section \ref{s.TFR_intro} gives a brief introduction to the Wigner-Ville Distribution, and explains why the interference terms which distort this time-frequency representation have resulted in the development of a large number of alternative representations. In section \ref{s.comparison} we
analyse the time-frequency resolution and statistical properties of six of these alternative time-frequency representations using three synthetic (noise-free) signals mimicking key features of
our data. We conclude that the Morlet wavelet scalogram, with the addition of wavelet ridges and maxima lines
is the optimal choice for our data, both in terms of its time-frequency resolution and statistical properties. We use this representation to re-analyse archival data of VW Hyi in section \ref{s.VWHyi}, and show that it does indeed provide a reliable tool for detecting QPOs, with a significantly higher detection rate than previous methods. Section \ref{s.conclusion} concludes.

\section{A brief introduction to time-frequency representations}
\label{s.TFR_intro}

\label{ss.wvd} The archetypal time-frequency representation (TFR) of
a signal $x(t)$ is the Wigner-Ville Distribution (WVD), given by

\begin{equation}
\label{eq.contWVD} W_x(t,f)=2\int_{-\infty}^{+\infty} x(t+\tau)
x^*(t-\tau)e^{4\pi f \tau i}d\tau
\end{equation}

\citep{kn.Vi48}.

In practise, for finite duration signals, the
pseudo Wigner-Ville distribution (PWVD), given by
\begin{equation}
\label{eq.contPWVD} W_x(t,f)=2\int_{-\infty}^{+\infty}
h(\tau)x(t+\tau) x^*(t-\tau)e^{4\pi f \tau i}d\tau.
\end{equation}
is used, where $h(t)$ is a rectangular window  that is zero outside
the duration of the signal. If $x(t)$ is sampled at
times $t=0,\Delta t, 2\Delta t, ...$ then, in order to
prevent spectral aliasing in the WVD and PWVD, either the signal must be
oversampled by a factor of at least 2, or the analytic signal \citep{kn.Au96}. The analytic version $x_a(t)$ of a
signal $x(t)$ is the inverse Fourier transform of
\begin{equation}
\hat{x}_a(f)=\left\{\begin{array}{ll}
2\hat{x}(f) & \mbox{if $f\geq0$}\\
0 & \mbox{if $f<0$}
\end{array}
\right.
\end{equation}

We assume throughout the rest of the paper that the analytic signal is used.

Ville originally investigated TFRs in order to study the instantaneous frequency of signals with time-varying components. A real (mono-component) signal $x(t)$ may
be written as a time varying amplitude $a(t)$ modulated by a time varying phase $\phi $:%
\begin{equation}
\label{eq.analytic}
x(t)=a(t)\cos [\phi (t)]
\end{equation}
with $a(t)\geq 0$. The instantaneous (angular) frequency is then
defined as the positive derivative of the phase:
\begin{equation}
\omega (t)=\phi ^{\prime }(t)\geq 0
\end{equation}
\citep{kn.Go97}. Since there are many possible choices of $a(t)$ and $\phi(t)$ for any given signal, $\omega(t)$ is not uniquely defined. However, using the analytic signal, we have $x_a(t)=a(t)\exp[i\phi(t)]$ and, since
$x(t)=\mathrm{Re}[x_a(t)]$, then $x(t)=a(t)\cos\phi(t)$. We call
$a(t)$ the analytic amplitude of $x(t)$ and $\phi ^{\prime }(t)$ its
instantaneous frequency \citep{kn.Ma98}. Multicomponent signals in which the components occur in disjoint frequency bands can be thought of as the sum of mono-component signals; it is the aim of the TFR to show the instantaneous frequency of each component accurately.

\begin{figure}
  \includegraphics[width=3.0in]{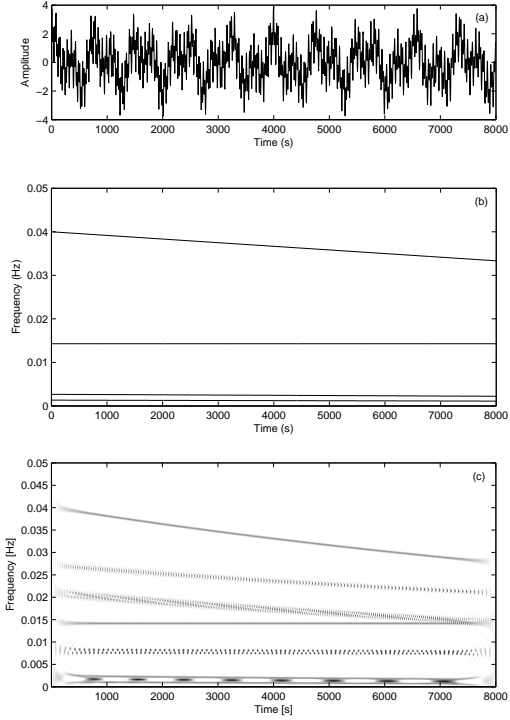}\\
  \caption{(a) Time-series of a synthetic
signal which mimics a noise-free dwarf nova (DN) lightcurve, with a
first harmonic DNO changing linearly in period from \un[25]{s} to
\un[30]{s}, fundamental and first harmonic QPOs starting with
periods \un[750]{s} and \un[375]{s} respectively and tracking the
period change of the DNO, and an lpDNO with a constant \un[70]{s}
period. (b) Idealised TFR of the signal in Figure 1(a). (c) WVD of
the signal in Figure 1(a)}\label{fig.ideal_tfr}
\end{figure}

Figure \ref{fig.ideal_tfr}(a) shows the time-series of a synthetic
signal which mimics a noise-free dwarf nova (DN) lightcurve sampled with an integration time of \un[4]{s}, with a
first harmonic DNO changing linearly in period from \un[25]{s} to
\un[30]{s} over the duration of the run, fundamental and first harmonic QPOs starting with
periods \un[750]{s} and \un[375]{s} respectively and tracking the
period change of the DNO, and an lpDNO with a constant \un[70]{s}
period, based on run s0484 of VW Hyi, discussed in \citet{kn.Wo02a}.
An idealised TFR (Figure \ref{fig.ideal_tfr}(b)) and the PWVD
(Figure \ref{fig.ideal_tfr}(c)) are shown below the time-series.

While the changing DNO and QPO periodicities are clearly visible in
the PWVD in Figure \ref{fig.ideal_tfr}(c), as is the constant period lpDNO, there are also five sets
of spurious signals, called interference terms (ITs), which are
an artefact of the quadratic nature of the WVD. The WVD of an N-component signal will always
consist of N signal terms and $\frac{N(N-1)}{2}$ ITs
\citep{kn.Me97}. ITs oscillate in the time and frequency directions
\citep{kn.Hl94}, and the closer together two components are in
frequency, the slower their ITs will oscillate in the time direction
\citep{kn.Hl92}. This is clear in figure \ref{fig.ideal_tfr}(c): the
ITs due to the interaction between the two close QPO terms oscillate
much more slowly in the time direction than the ITs due to the other
interactions, whose generating signal terms are further apart in
frequency.

The presence of ITs in the WVD make it unsuitable for analysis of
data in which the true signal components are unknown, or cannot be
inferred by other means - as is the case for our data. However,
since the ITs are oscillatory, they can be reduced or removed
entirely by smoothing the WVD in the time and frequency directions.
IT attenuation is usually achieved by convolving the WVD with a 2-D
smoothing kernel \citep{kn.Hl92}.

IT reduction comes at a cost: the more smoothing in time
and/or frequency, the poorer the resolution in time and/or frequency
\citep{kn.Co89}. The WVD has the best possible time-frequency
resolution of all TFRs: for a time-dirac signal centered at $t_0$,
$W_x(t,f)=\delta(t-t_0)$ , and for a frequency dirac signal centered
at $f_0$, $W_x(t,f)=\delta(f-f_0)$ \citep{kn.Ma98}. All other TFRs
generally have poorer time and frequency resolution, or may achieve
that of the WVD for certain types of signal only.

Since the geometry of the ITs depends on the components of the
signal, different smoothing kernels have been developed for
different kinds of signals: there is no single kernel which is
optimal for all signals. Most kernels however belong to the Cohen
class \citep{kn.Co89} or the affine class \citep{kn.Ri92} or both,
and have one or more parameters which can be varied to choose the
amount of smoothing in time and/or frequency.

The Cohen TFRs \citep{kn.Co89}, which we will denote generally as
$C_x(t,f)$, are derived from a time-frequency convolution of the WVD
with a smoothing kernel $\psi(t,f)$:

\begin{equation}
\label{eq.cohen_tf}
C_x(t,f)=\int_{t'}\int_{f'}\psi(t-t',f-f')W_x(t',f')dt'df'
\end{equation}

or, more traditionally, as
\begin{equation}
\label{eq.cohen_doppler}
C_x(t,f)=\int_{\tau}\int_{\nu}\Psi(\tau,\nu)\Lambda_x(\tau,\nu)e^{i2\pi
(t\nu -f\tau)}d\tau d\nu
\end{equation}

where $\Psi(\tau,\nu)$ is the Fourier transform of $\psi(t,f)$ and
$\Lambda_x(\tau,\nu)$ is the ambiguity function (AF) of $x(t)$, defined
as:

\begin{equation}
\begin{array}{ll}
\Lambda_x(\tau,\nu) & =2\int_t
x(t+\tau)x^*(t-\tau)e^{-i4\pi \nu t}dt\\
&=2\int_f\hat{x}(f+\nu) \hat{x}^*(f-\nu) e^{i4\pi \tau f}df.\\
\end{array}
\end{equation}

Here $\tau$ is the time-delay, and $\nu$ is the Doppler value, which
measures the difference in frequencies between points at a given
time-delay \citep{kn.Hl95}. The AF, which is the Fourier transform
of the WVD, can be interpreted as a joint time-frequency correlation
function \citep{kn.Hl92}.

Each member of Cohen's class is thus associated with a unique,
signal independent kernel function $\Psi(\tau,\nu)$; the WVD is a
member of the Cohen class, with kernel $\Psi_{WVD}(\tau,\nu)=1$.

The general affine TFR ($A_x(t,a;\Pi)$) of an analytic signal $x(t)$
is given by

\beq
A_x(t,a;\Pi)=\int_{-\infty}^{\infty}\int_{-\infty}^{\infty}\Pi(\frac{s-t}{a},a\xi)W_x(s,\xi)ds
d\xi \eeq

where $\Pi(t,\nu)$ is a smoothing kernel. Equivalently, using the AF
instead of the WVD, we have

\beq
\label{eq.affineTFR}
A_x(t,a;\Psi)=\int_{-\infty}^{\infty}\int_{-\infty}^{\infty}\Psi(a\xi,\tau/a)\Lambda_x(\xi,\tau)e^{-i2\pi
\xi t}d\xi d\tau \eeq

where $\Psi(\xi, \tau)$ is the ambiguity domain version of
$\Pi(t,\nu)$. Affine TFRs are time-\emph{scale}
representations rather than time-frequency representations, but $a$
acts as a measure of the frequency: reducing $a$  reduces the time
support of $\Pi_{t,s}$, and hence increases its frequency range
\citep{kn.Go99}. Since the WVD is scale-invariant, it is also a
member of the affine class.

In both Cohen and affine TFRs, the kernel determines the amount of
smoothing in time and/or frequency. The important difference between
affine and Cohen TFRs is that the affine kernel is scaled depending
on the analysing frequency, while Cohen kernels retain the same
`size' at all frequencies. This means that for affine TFRs the time
and frequency resolutions depend on the frequency at which the
signal is being analysed, while for Cohen TFRs the time and
frequency resolutions are the same at all frequencies.

\begin{deluxetable}{p{3.5cm}ccccc}
\tablecaption{TFR kernels and their properties\label{tbl.kernels}}
\tabletypesize{\small}
\tablehead{\colhead{Name} & \colhead{Kernel $\Psi(\tau,\nu)$} & \colhead{Parameter(s)} & \colhead{Cohen} & \colhead{Affine} & \colhead{Statistics}}
\startdata
Wigner-Ville (WVD) & 1 & - & \checkmark & \checkmark & $S_{noise}(f)\frac{\chi_2^2}{2}$\\
&& & & &\\
Smoothed Pseudo Wigner-Ville (SPWVD) & $g(\frac{\tau}{2})g^*(-\frac{\tau}{2})\hat{h}(\nu)$ & $L_g$, $L_h$ & \checkmark & &$\frac{2}{2M-1}S_{noise}(f)\frac{\chi_2^2}{2}$\\
&& & & &\\
Choi-Williams (CWD) & $e^{-\frac{(2\pi \tau \nu)^2}{\sigma}}$ & $\sigma$ & \checkmark & & $S_{noise}(f)\frac{\chi_2^2}{2}$\\
&& & & &\\
Born-Jordan (BJD) & $\frac{\sin(\pi \tau \nu)}{\pi \tau \nu}$ & - & \checkmark & & $S_{noise}(f)\frac{\chi_2^2}{2}$\\
&& & & &\\
Cone-Kernel (CKR) & $g(\tau) |\tau|\frac{\sin(\pi \tau \nu)}{\pi \tau \nu}$ & $L_g$ & \checkmark & & $S_{noise}(f)\frac{\chi_2^2}{2}$\\
&& & & &\\
Affine Smoothed Pseudo Wigner-Ville (ASPWVD) & $h(\frac{\tau}{a})g(\frac{s-\tau}{a})$ & $L_g$, $L_h$ & & \checkmark & $\frac{2}{2M_s-1}S_{noise}(s)\frac{\chi_2^2}{2}$ \\
&& & & &\\
Gabor Spectrogram (SPEC) & $W_h(-\tau, -\nu)$ & $L_h$ & \checkmark & & $S_{noise}(f)\frac{\chi_2^2}{2}$\\
&& & & &\\
Scalogram (SCALO) & $W_{\psi}(\tau, \nu)$ & $L_{\psi}$ & & \checkmark & $S_{noise}(f)\frac{\chi_2^2}{2}$\\
  \hline
\enddata
\end{deluxetable}

\section{Six smoothing kernels investigated}
\label{s.comparison}

We will investigate 4 Cohen and 2 affine smoothing kernels which have either been previously used in used in astronomy or have properties which suggest they are appropriate for our data; Table \ref{tbl.kernels}
gives a summary of the main features of the kernels discussed. We begin our analysis by finding the optimal parameters needed for each TFR in order to represent the synthetic DN
lightcurve introduced in section \ref{ss.wvd} without interference terms. This signal tests the
frequency resolution of each TFR over a broad frequency range, and ensures that
the values of the parameters chosen completely attenuate
interference terms over the whole frequency range. Using these
optimal parameters, we then further investigate the time resolution of each
TFR by using intermittent sinusoids of different frequencies. Where possible, we include a discussion the use of the kernel in the astronomical literature. We conclude this section with a discussion of the statistical properties of the six kernels.

\subsection{The Smoothed Pseudo WVD}
\label{ss.spwvd}

One method of smoothing the WVD is to use the separable kernel

\begin{equation}
\Psi_{SPWVD}(\tau,\nu)=g(\frac{\tau}{2})g^*(-\frac{\tau}{2})\hat{h}(-\nu)
\end{equation}

with time-smoothing window $g(t)$ and frequency-smoothing window
$\hat{h}(\nu)$ (which is the Fourier transform of a time-smoothing
window $h(t)$) \citep{kn.Au96}. The length $L_g$ ($L_h$) of the
window $g(t)$ ($h(t)$) controls the amount of smoothing in the time
(frequency) domain. A longer $g(t)$ gives more time smoothing and
hence worse time resolution, and a longer $h(t)$ gives less
frequency smoothing and hence better frequency
resolution \citep{kn.Hl95}.

\begin{figure}
  \includegraphics[width=3in]{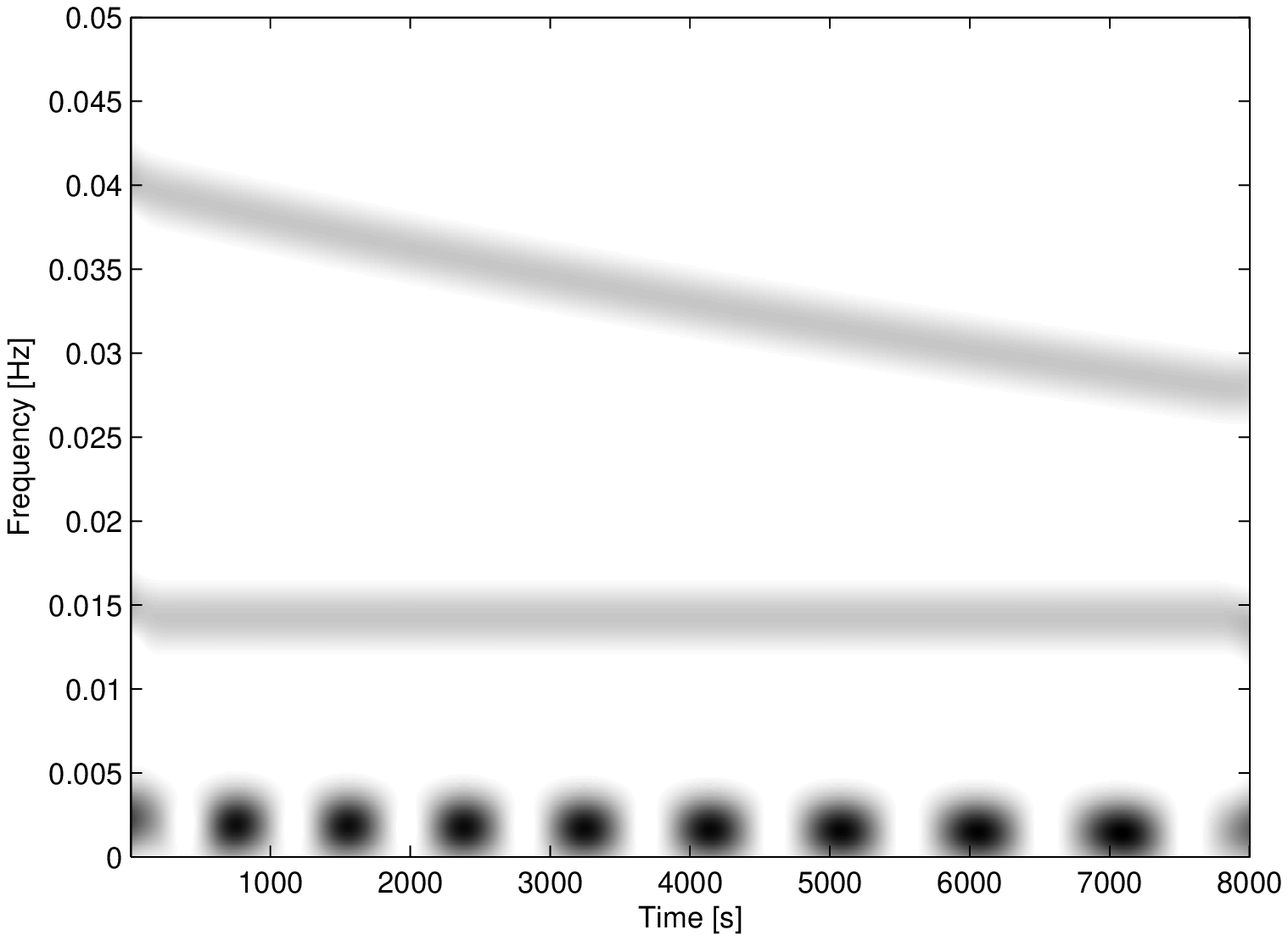}\\
  \caption{Smoothed Pseudo Wigner-Ville distribution of synthetic DN lightcurve, with L$_g=$200s, L$_h=$200.}\label{fig.spwvd_sig8_g101_h101}
\end{figure}

\begin{figure}
  \includegraphics[width=3in]{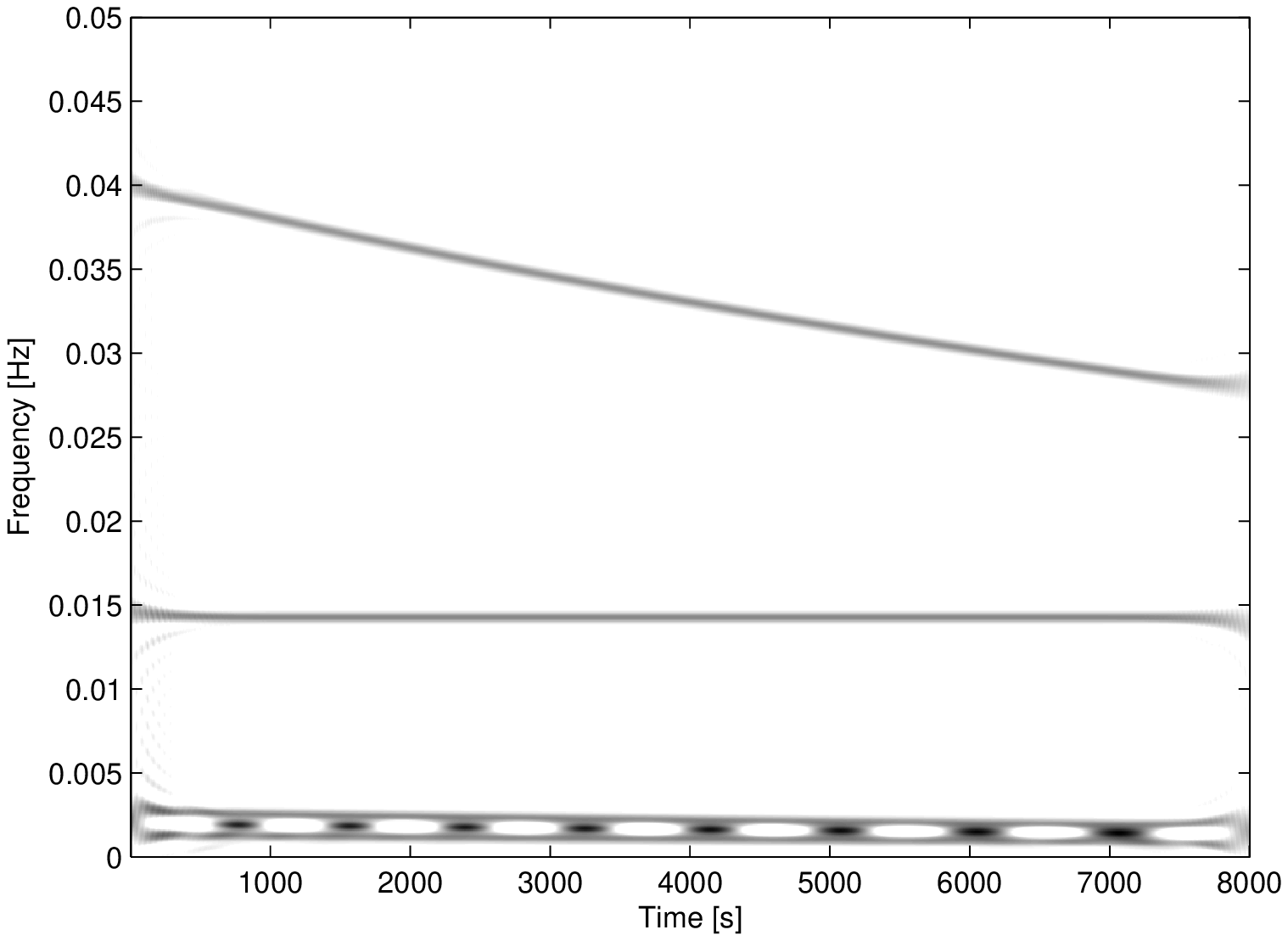}\\
  \caption{Smoothed Pseudo Wigner-Ville distribution of the synthetic DN lightcurve, with L$_g=$200s, L$_h=$1200s.}\label{fig.spwvd_sig8_g101_h601}
\end{figure}

\begin{figure}
  \includegraphics[width=3in]{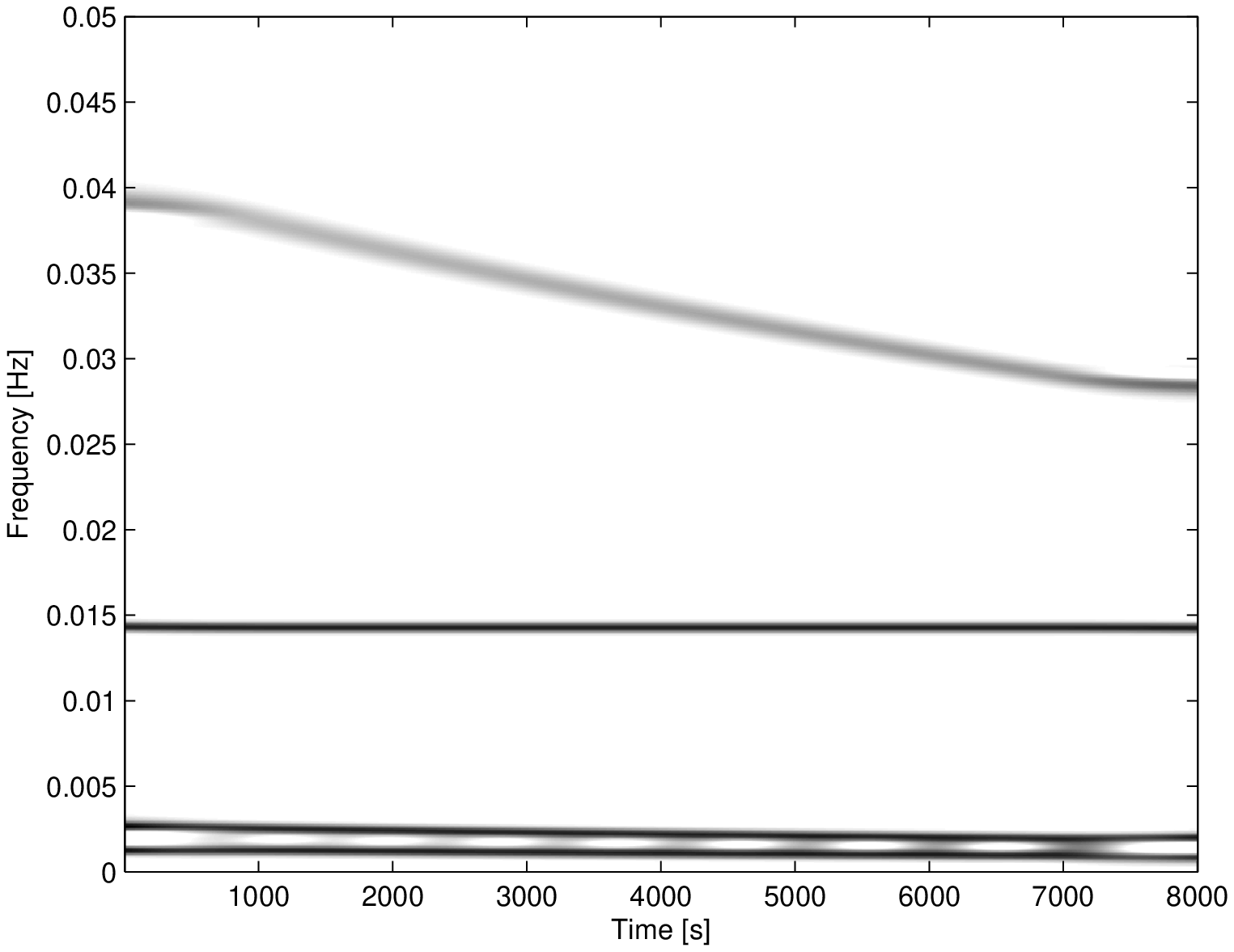}\\
  \caption{Smoothed Pseudo Wigner-Ville distribution of the synthetic DN lightcurve, with L$_g=$1000s, L$_h=$1200s.}\label{fig.spwvd_sig8_g501_h601}
\end{figure}

Figures \ref{fig.spwvd_sig8_g101_h101} and
\ref{fig.spwvd_sig8_g101_h601} show the smoothed pseudo WVD (SPWVD)
of the synthetic DN lightcurve, with $L_g=$ \un[200]{s} in both
figures, but $L_h=$ \un[200]{s} in the first, and $L_h=$
\un[1200]{s} in the second. The effect of increasing $L_h$ (i.e.
decreasing the frequency smoothing) is marked: in the first figure,
we cannot differentiate the two long period QPO components, but in
the second they are clearly visible. Interference terms between the
two long-period components in Figure \ref{fig.spwvd_sig8_g101_h601}
are still prominent. By increasing $L_g$ to \un[1000]{s}, as in
figure \ref{fig.spwvd_sig8_g501_h601}, the interference terms are
completely attenuated. Notice, however, the effect of the extensive
time smoothing at the ends of the DNO.

\begin{figure}
  \includegraphics[width=3in]{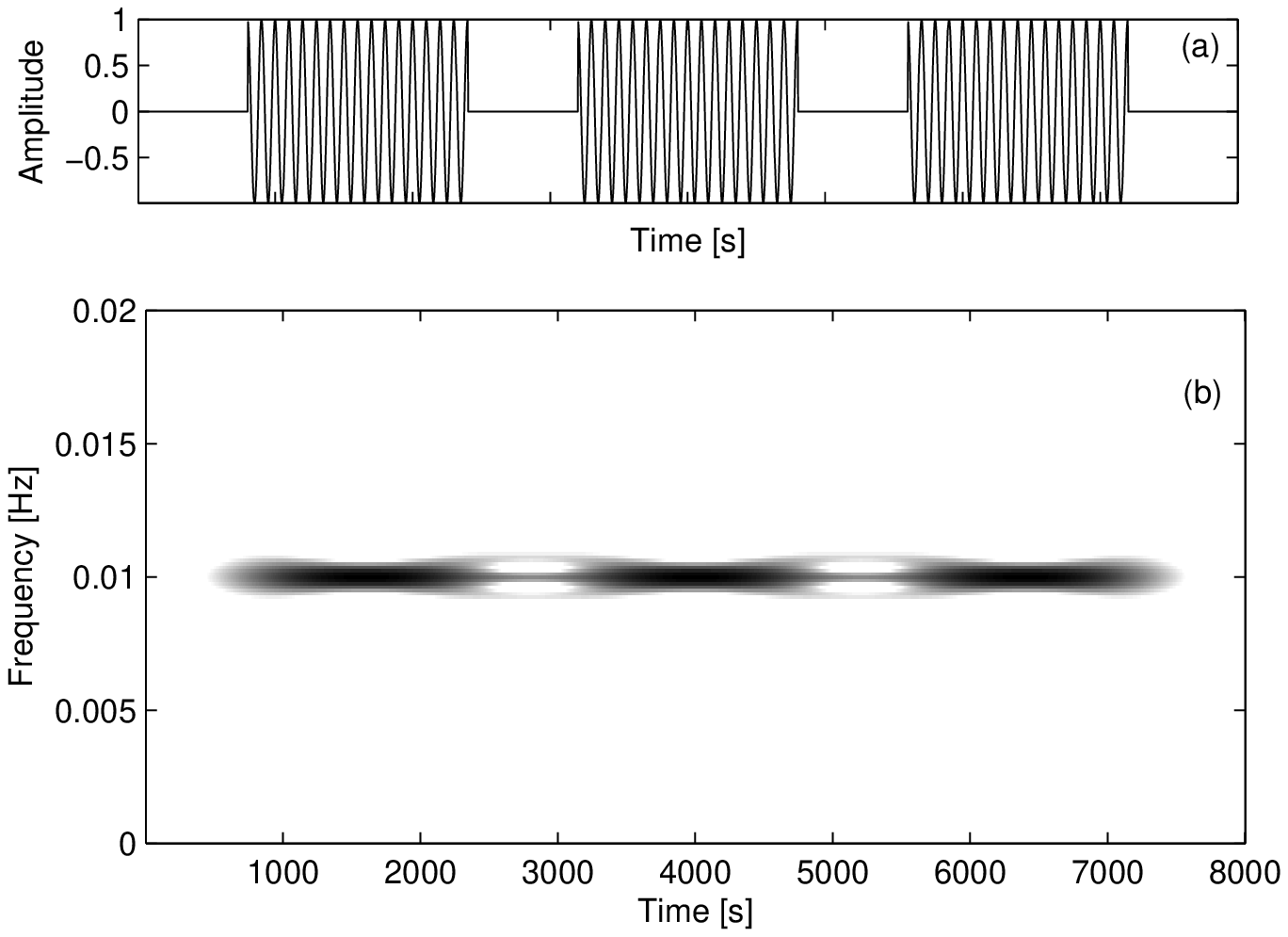}\\
  \caption{Smoothed Pseudo Wigner-Ville distribution of an intermittent sinusoid of period \un[100]{s}, with L$_g$ 1000s, L$_h$ 1200s.}\label{fig.spwvd_sig12_g101_h601}
\end{figure}

Next, we use these optimal values of $L_g$ and $L_h$ to investigate an intermittent sinusoid of period \un[100]{s}, shown figure
\ref{fig.spwvd_sig12_g101_h601}(a), mimicking for example an intermittent lpDNO. The integration time is \un[4]{s}.

The time-smoothing introduced by increasing $L_g$, which was
necessary to attenuate the ITs, has come at a price: the
ability of the SPWVD to clearly resolve temporal discontinuities in
the signal. In figure \ref{fig.spwvd_sig12_g101_h601}(b) the
smoothing in the time domain results in a SPWVD that does not
clearly show the beginning or end of each sinusoidal burst. Indeed,
it gives the impression that the sinusoid is perhaps present for the
duration of the signal, but at varying amplitude. There are two
reasons for this: signal terms, which should not be smoothed, are
smeared in the time dimension by the long time window, and ITs which
oscillate in the frequency direction between the signal terms are
not removed because we have set the length of the frequency
smoothing window to ensure that we have optimal frequency resolution
(i.e. minimal smoothing).

We are thus presented with the central issue that dogs most Cohen
kernels: if ITs are attenuated by smoothing in time, the result is
poor time resolution, which is particularly problematic at high
frequencies.

\subsection{The Affine Smoothed Pseudo WVD}
\label{ss.aspwvd}

The affine smoothed pseudo WVD (ASPWVD) is the affine counterpart of
the SPWVD, and its kernel is given by

\beq \Psi_{ASPWVD}=h(\frac{\tau}{a})g(\frac{s-t}{a}),\eeq

where $g(t)$ and $h(t)$ are smoothing windows as for the SPWVD.
The lengths of the time and frequency smoothing windows, $L_g$ and $L_h$ respectively, again allow a flexible choice of time and scale
(frequency) resolution.

\begin{figure}
  \includegraphics[width=3in]{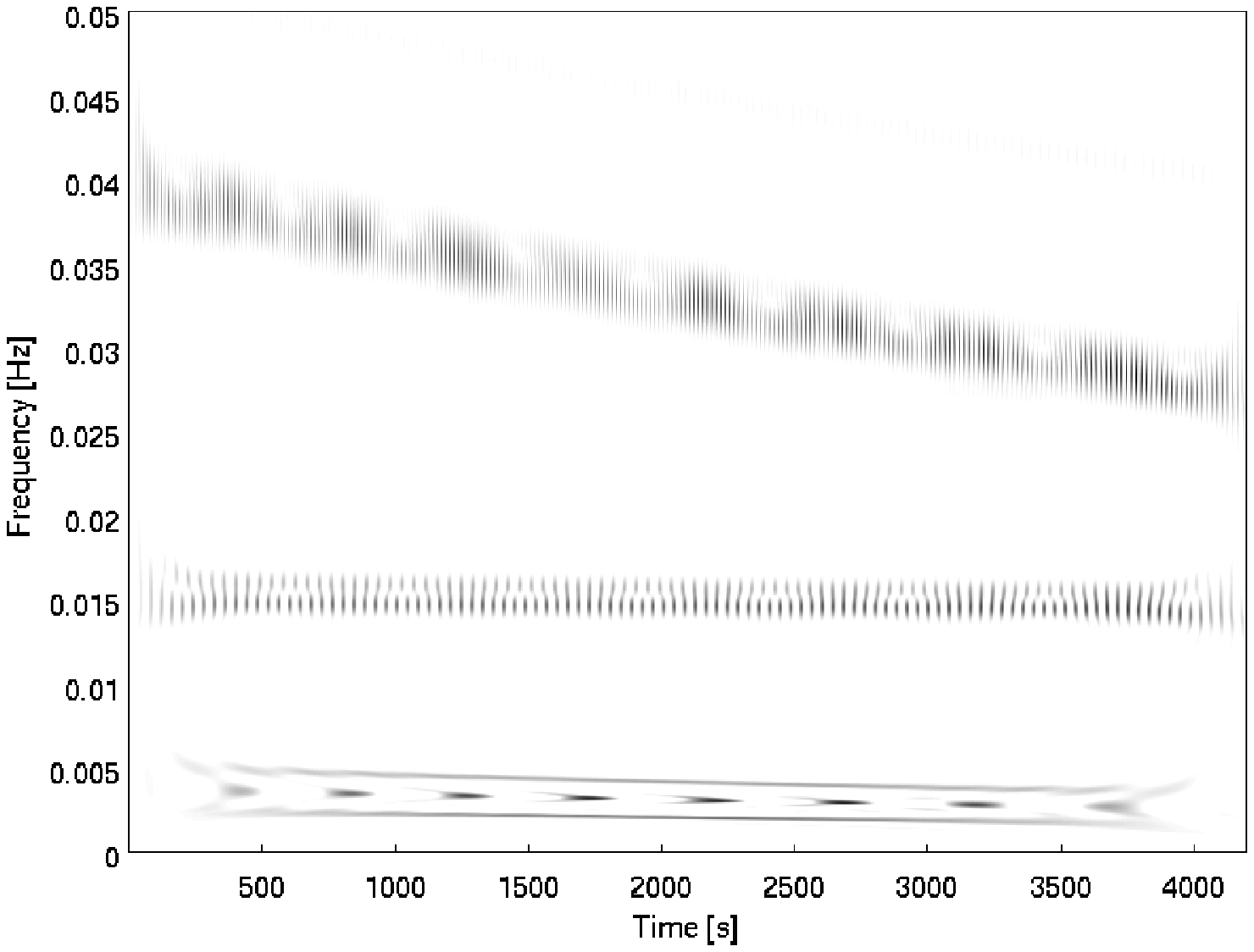}\\
  \caption{ASPWVD of the synthetic DN lightcurve, with $L_g=0$s and $L_h=320s.$}\label{fig.spawvd_sig8_g0_h80}
\end{figure}

\begin{figure}
  \includegraphics[width=3in]{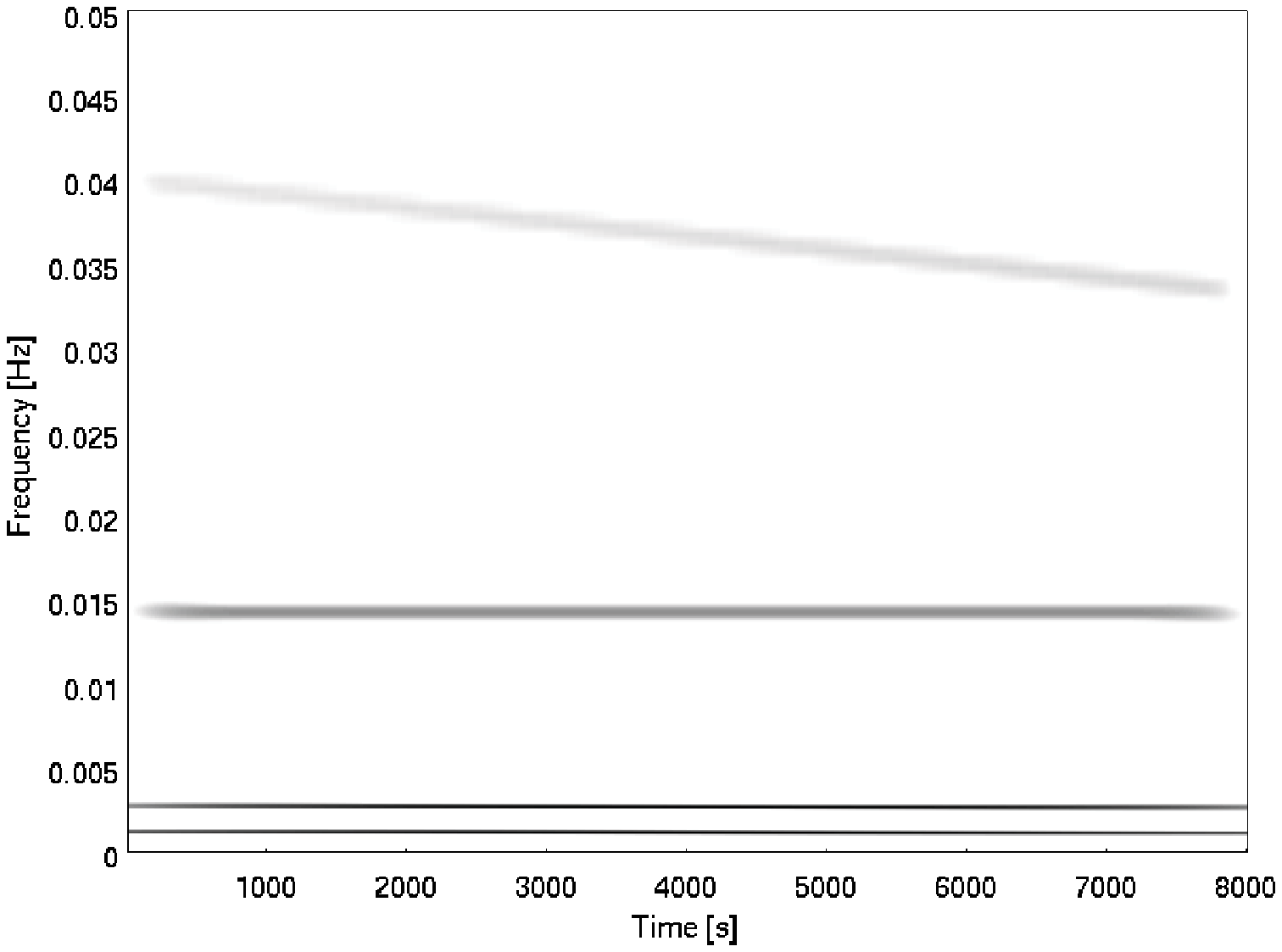}\\
  \caption{ASPWVD of the synthetic DN lightcurve, with $L_g=320$s and $L_h=320$s.}\label{fig.spawvd_sig8_g80_h80}
\end{figure}

Figures \ref{fig.spawvd_sig8_g0_h80} and
\ref{fig.spawvd_sig8_g80_h80} show the ASPWVD of the synthetic DN
lightcurve, with $L_g$ set to \un[0]{s} in the first case (no time
smoothing) and \un[320]{s} (at the largest scale) in the second.
$L_h$ is \un[320]{s} in both cases, which is the minimum length
needed to resolve both QPOs. What is apparent in Figure
\ref{fig.spawvd_sig8_g80_h80}, when compared with the SPWVD shown in
Figure \ref{fig.spwvd_sig8_g501_h601}, is that less time smoothing
is needed to remove the ITs in the ASPWVD, resulting in less
noticeable end-effects. However, the frequency resolution decreases
in the ASPWVD as the frequency increases, so the DNO in the ASPWVD
is broader than the DNO in the PSWVD.

\begin{figure}
  \includegraphics[width=3in]{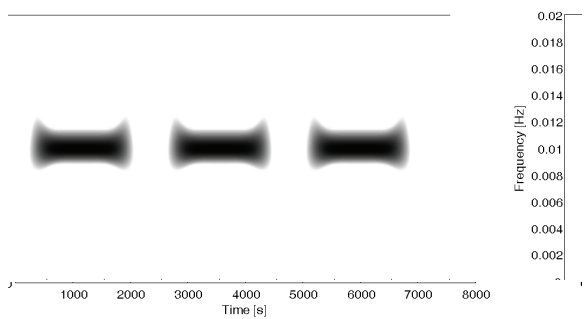}\\
  \caption{ASPWVD of an intermittent sinusoid of period \un[100]{s}, with $L_g=320$s and $L_h=320$s.}\label{fig.spawvd_sig6_g80_h80}
\end{figure}

In comparison with the SPWVD, the ASPWVD of the intermittent sinusoid, shown in figure
\ref{fig.spawvd_sig6_g80_h80}, has good time resolution (since for
affine distributions the time resolution improves as the frequency
increases), but has poorer frequency resolution than the SPWVD. No ITs are visible between the bursts of signal.

\begin{figure}
  \includegraphics[width=3in]{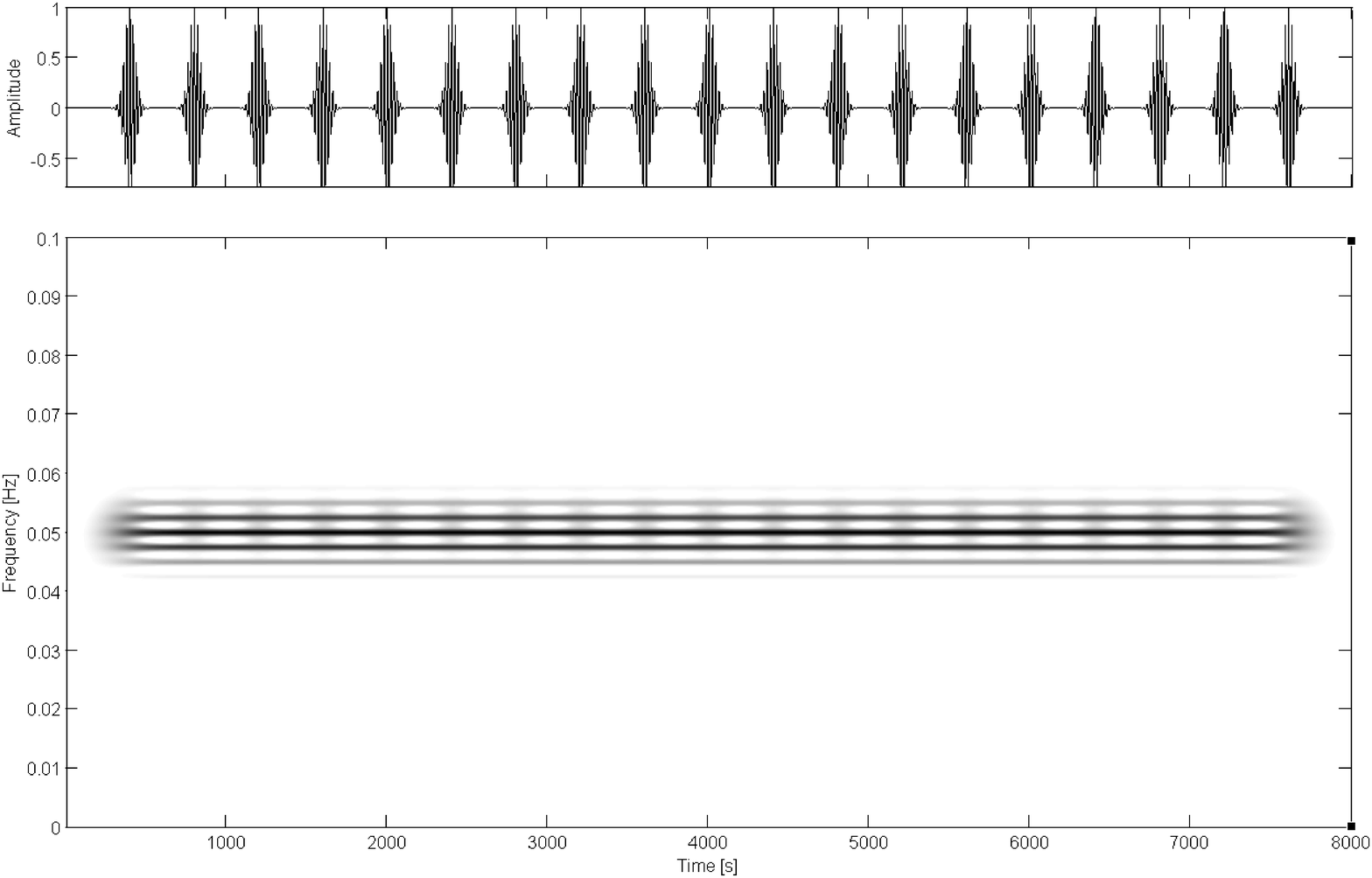}\\
  \caption{ASPWVD of an intermittent, amplitude modulated sinusoid of period \un[20]{s}, with $L_g=320$s and $L_h=320$s.}\label{fig.spawvd_burst_g80_h80}
\end{figure}

The top panel of figure \ref{fig.spawvd_burst_g80_h80} shows a signal mimicking an intermittent \un[20]{s} DNO with changing amplitude, sampled with an integration time of \un[4]{s}. The ASPWVD of this signal, calculated using $L_g=L_h=$ \un[320]{s} is shown in the bottom panel. Very broadly smoothed signal terms and ITs considerably distort the true signal, and it is not possible to remove these interference terms without losing the optimal resolution achieved for the analysis of the DN signal.

\subsection{The Choi-Williams Distribution}
\label{ss.cwd}

Instead of using a separable smoothing kernel, several kernels have
been developed that depend on the product of $\tau$ and $\nu$. One
of the earliest examples is the Choi-Williams distribution (CWD)
\citep{kn.Ch89}, which is characterized by the kernel

\begin{equation}
\Psi_{CWD}(\tau, \nu)=e^{-\frac{(2\pi \tau \nu)^2}{\sigma}}.
\end{equation}

$\sigma$ is a positive parameter that jointly controls smoothing
in time and frequency: a larger value for $\sigma$ gives less
smoothing \citep{kn.Hl95}. In fact, as $\sigma \rightarrow +\infty$,
the CWD tends to the WVD. Pure sinusoids have the ideal
time-frequency resolution of the WVD, but all other signals are
broadened when compared with the WVD \citep{kn.Hl94}.

\begin{figure}
  \includegraphics[width=3in]{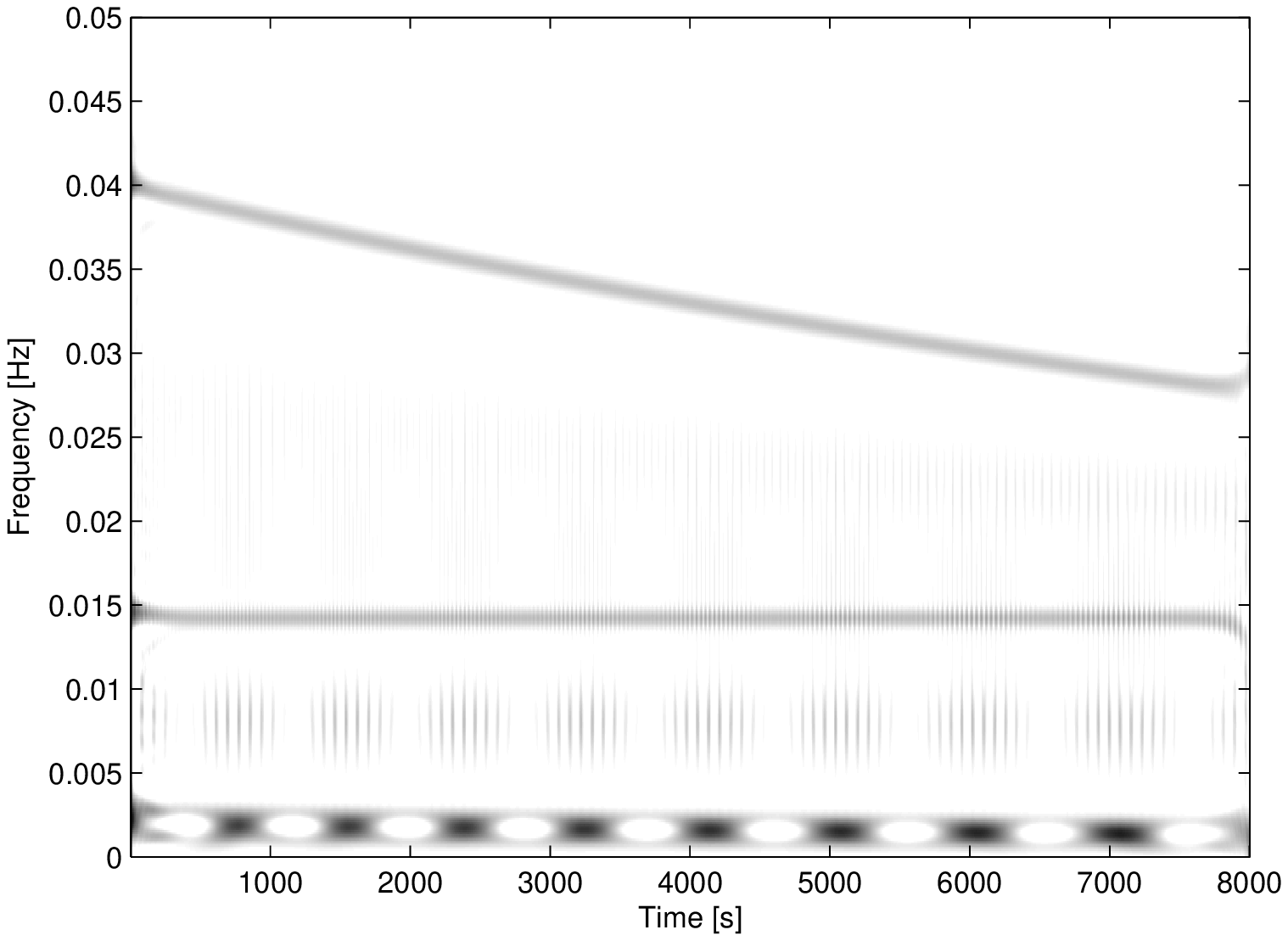}\\
  \caption{CWD of the synthetic DN lightcurve with $\sigma = 100$.}\label{fig.cw_sig8_sigma_100}
\end{figure}

\begin{figure}
  \includegraphics[width=3in]{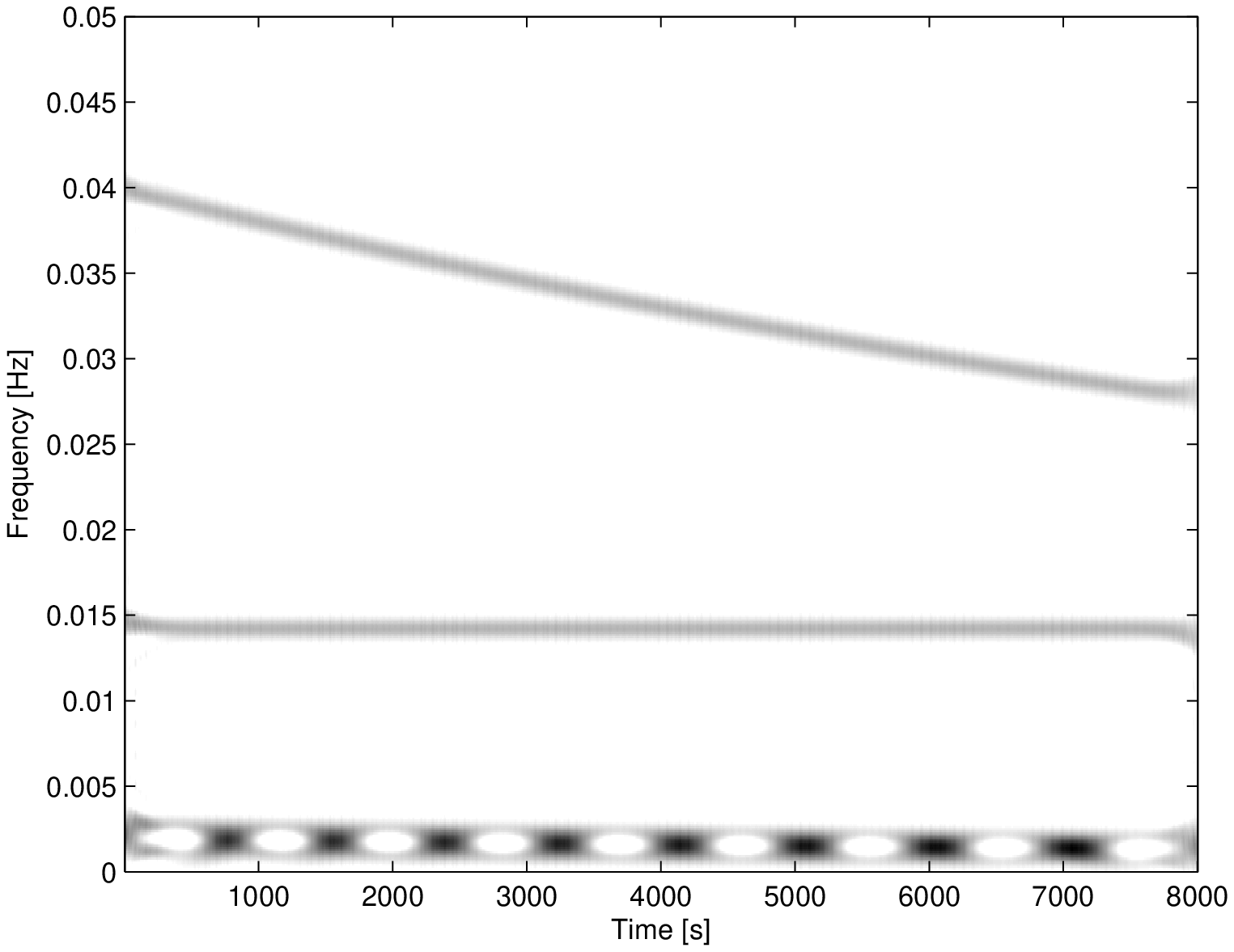}\\
  \caption{CWD of the synthetic DN lightcurve with $\sigma = 0.1$.}\label{fig.cw_sig8_sigma_01}
\end{figure}

Figure \ref{fig.cw_sig8_sigma_100} shows the CWD of the synthetic DN
lightcurve with $\sigma=100$. Reducing $\sigma$ to 0.1, as shown in
figure \ref{fig.cw_sig8_sigma_01}, which increases the amount of
smoothing, does not fully eliminate the ITs; further reductions in
$\sigma$ have no noticeable affect. In fact, ITs due to
different frequency components occurring at the same time cannot be
attenuated completely if we use the CWD \citep{kn.Hl95}. In \citet{kn.Ki02a}, the good time
resolution of the CWD enables the increase in frequency of the main
pulsation to be seen clearly, and the strong interference terms caused
by the interaction of the main component with its harmonics are not
a problem, since they can be easily identified as spurious using the
periodogram. However, for our data, since complete IT attenuation is required, the CWD is not ideal.

\subsection{The Born-Jordan Distribution and Cone-shaped Kernel Representation}
\label{ss.bjd}

If we require a TFR that preserves finite time- and
frequency-support, then the simplest choice for $\Psi(\tau,\nu)$ is

\begin{equation}
\Psi_{BJ}(\tau,\nu)=\frac{\sin(\pi \tau \nu)}{\pi \tau \nu},
\end{equation}

which defines the Born-Jordan distribution (BJD). This kernel does
not admit smoothing of ITs in the frequency direction. Allowing
smoothing of the BJD in frequency gives the Zhao-Atlas-Marks
representation \citep{kn.Zh90}, also known as the cone-shaped kernel
representation (CKR), with the kernel

\begin{equation}
\Psi_{CKR}(\tau,\nu)=g(\tau)|\tau|\frac{\sin(\pi \tau \nu)}{\pi \tau
\nu}
\end{equation}

where $g(\tau)$ is any window function \citep{kn.Hl95}. The length of the window, $L_g$, again
controls the amount of frequency smoothing.

\begin{figure}
  \includegraphics[width=3in]{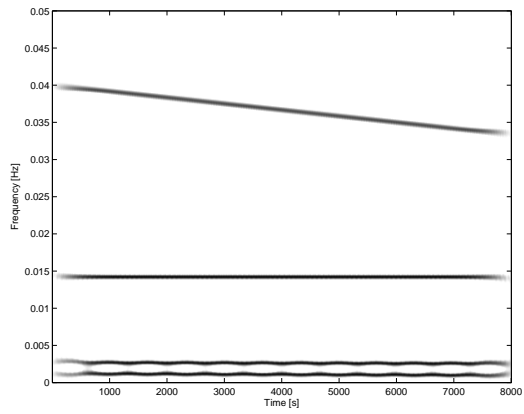}\\
  \caption{CKR of the synthetic DN lightcurve with $L_g=1000$s.}\label{fig.ckr_dn_g501_h501}
\end{figure}

Figure \ref{fig.ckr_dn_g501_h501} shows the CKR of the synthetic DN signal
with $L_g=$ \un[1000]{s}, which is the minimal length required to
completely attenuate the ITs between the QPOs. The frequency resolution is comparable to the SPWVD and ASPWVD, and the edge effects are less marked than those of the ASPWVD. Notice, however, that the way in which the CKR smooths the signal in figure \ref{fig.ckr_dn_g501_h501} causes the QPOs to appear to
oscillate slightly in frequency. IT attenuation in the
CKR is complex. For components which are close in frequency, the
attenuation will depend on the time distance, $\tau$, between the components.
If the time distance between two components is $\tau$, and
$0<|\tau|<\frac{1}{2}L_g$ then the resulting IT will be amplified,
rather than attenuated. However all terms (signal and IT) with
$|\tau|>\frac{1}{2}L_g$ will be completely suppressed. We are thus
faced with a dilemma: increasing $L_g$ improves frequency
resolution, but also makes it more likely that ITs will fall in the
$0<|\tau|<\frac{1}{2}L_g$ range, and hence be enhanced
\citep{kn.Hl95}.

\begin{figure}
  \includegraphics[width=3in]{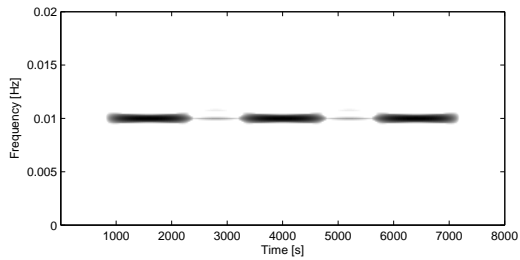}\\
  \caption{CKR of a intermittent sinusoid of period \un[100]{s}, L$_g$=1000s.}\label{fig.zam_sig6_g501_h501}
\end{figure}

The CKR of the intermittent \un[100]{s} is shown in figure \ref{fig.zam_sig6_g501_h501}. In comparison with the previous two TFRs, the frequency resolution of the CKR is excellent. The time resolution is similar to that of the ASPWVD. ITs, although of low amplitude, are present, again appearing as a low amplitude \un[100]{s} oscillation between the true signal bursts.

\begin{figure}
  \includegraphics[width=3in]{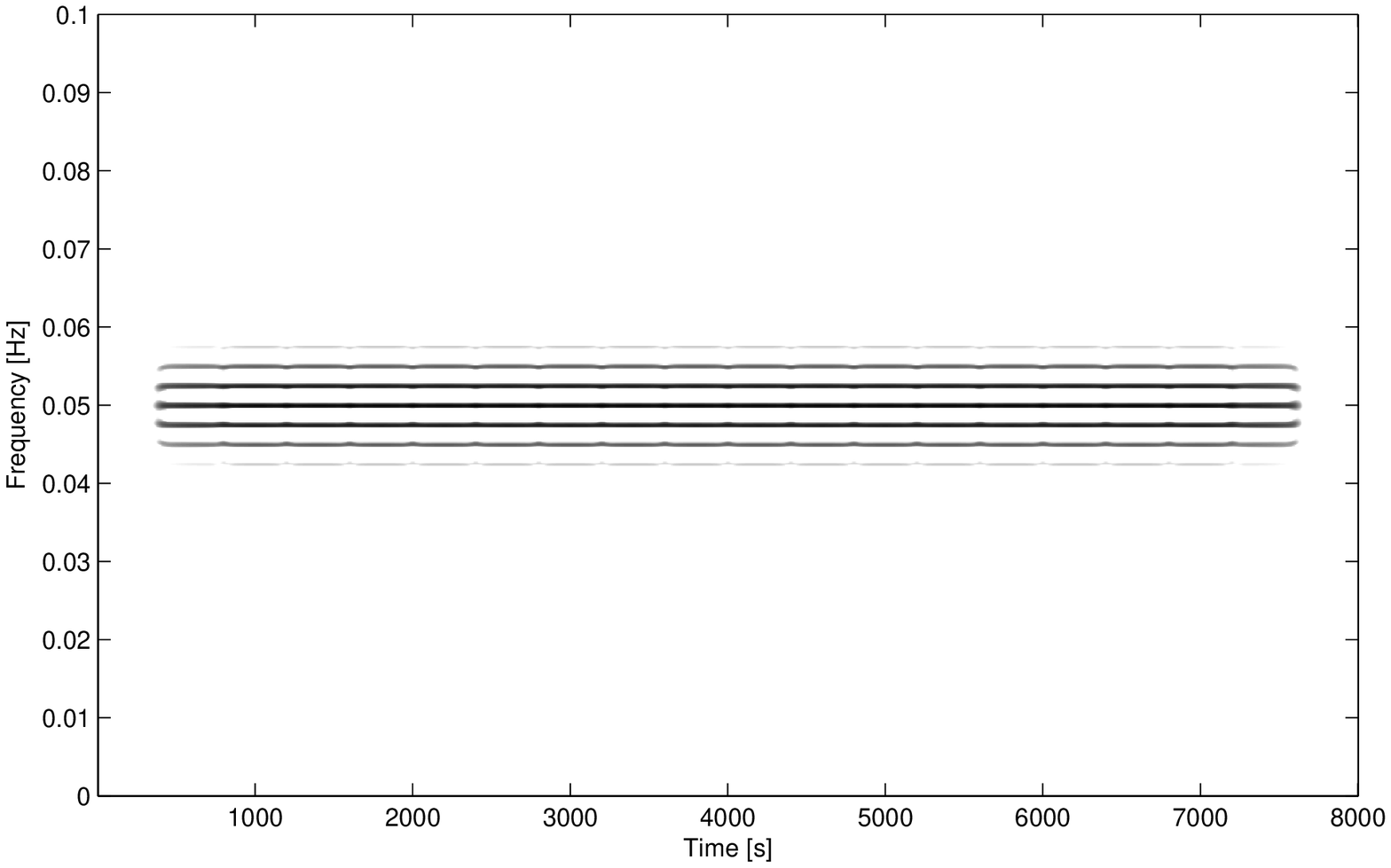}\\
  \caption{CKR of an intermittent, amplitude modulated sinusoid of period \un[20]{s}, with $L_g=1000$s and $L_h=1000$s.}\label{fig.ckr_burst_g501_h501}
\end{figure}

Figure \ref{fig.ckr_burst_g501_h501} shows the CKR of the intermittent, amplitude-modulated \un[20]{s} oscillation signal.
The amount of smoothing in time required to remove the low-frequency
ITs in the synthetic DN lightcurve results in the signal terms of the
intermittent DNO being completely smoothed. In addition, the ITs are
amplified (due to the complicated IT attenuation of the
CKR kernel) and smoothed , resulting in the appearance of five continuous
frequency bands, even more pronounced than those in the ASPWVD of the same signal. While this is a somewhat extreme example, it does
show that the effects of the CKR must be carefully monitored.

As the CKR can attenuate signals in a manner that
is not easy to predict, and can amplify ITs rather than attenuating
them, it is not ideal for analyses requiring significance testing. For large amplitude signals where this is not the case, the CKR can be an extremely useful tool: in figures 2 to 4
of \citet{kn.Bu04}, for example, the CKR is seen to give sharp images for the
five large-amplitude irregular pulsators under discussion. The
fundamental period and harmonics are visible; and since these are
relatively widely spaced in frequency, the ITs are easily
identifiable.

\subsection{Gabor Spectrogram}
\label{ss.gabor}

The (Gabor) spectrogram's kernel is given by

\begin{equation}
\Psi_{SPEC}=W_h(-\tau,-\nu),
\end{equation}

where $W_h(-\tau,-\nu)$ is the WVD of $h(t)$, any short-time
analysis window localised around $t=0$ and $f=0$. This means that
the time-frequency resolution of the spectrogram is dictated by the
resolution of $h(t)$, rather than that of the analysed signal.
Smoothing in the spectrogram is thus extensive, resulting in
virtually complete interference attenuation, but also in poor
time-frequency resolution \citep{kn.Hl92}. The choice of $h(t)$
determines the trade-off between the time spread and the frequency
spread of the smoothing function: time resolution is proportional to
the time duration of $h(t)$, while frequency resolution is
proportional to the bandwidth of $h(t)$.

\begin{figure}
  \includegraphics[width=3in]{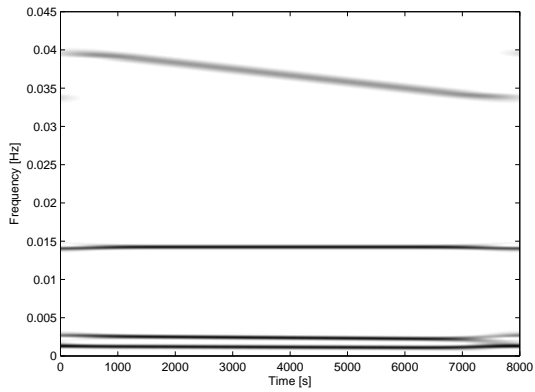}\\
  \caption{Gabor spectrogram of the synthetic DN lightcurve, with $L_g=2000$s.}\label{fig.gabor_sig8}
\end{figure}

Figure \ref{fig.gabor_sig8} shows the Gabor spectrogram of our
synthetic DN signal, using a Gaussian window of length \un[2000]{s},
which is the shortest that allows resolution of the QPO components. Each
component is clearly visible, and there are no interference terms,
although some smoothing is apparent near the ends.

\begin{figure}
  \includegraphics[width=3in]{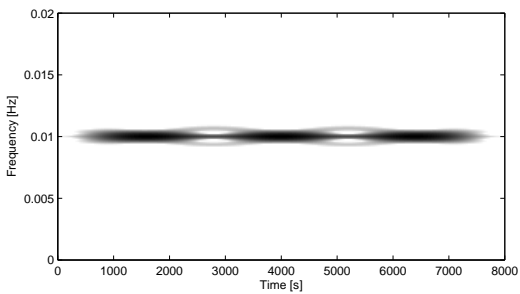}\\
  \caption{Gabor spectrogram of an intermittent sinusoid, with window length 2000s.}\label{fig.gabor_sig6}
\end{figure}

Figure \ref{fig.gabor_sig6} shows the Gabor spectrogram of an
intermittent sinusoid, using the same window as in the previous
example. We see that requiring good frequency resolution has again
resulted in poor time resolution. Indeed, as \citet{kn.Co89} and
\citet{kn.Zh90} point out, if one wishes to have accurate time and
frequency measurements using the spectrogram, two separate analyses
must be undertaken, using windows of different length.

\citet{kn.Bo95} use the Gabor transform to investigate HD 60435 (see
fig. 7 of \citet{kn.Bo95}). Because they are analysing a specific
oscillation (11.64 min) they are able to choose an analysis window
width which gives the required time-frequency resolution in this
narrow frequency band. \citet{kn.Be04} also consider a relatively
narrow frequency band, and are hence able to achieve an appropriate
time-frequency resolution balance. \citet{kn.Do98} gives one of the
the few examples of the analysis of a transient signal with the
Gabor spectrogram, but since the frequency range is relatively
narrow, they are able to choose a window length which gives the
optimal time-frequency resolution for analyzing the signal.

\subsection{The Wavelet Scalogram}
\label{sss.tfr_wavelet}

The wavelet scalogram is the affine counterpart of the Gabor spectrogram. Just as the Gabor spectrogram can be calculated using any short-time analysis window that is localised around $t=0$ and $f=0$, the wavelet scalogram can be calculated using any ``mother wavelet'' $\psi _{0}$ that is centered in the neighbourhood of $t=0$, has zero average ($\int^{\infty}_{\infty}\psi_0(t)dt=0$) and is normalized to have unit energy ($\int^{\infty}_{\infty}|\psi_0(\omega)|^2 d\omega=1$) \citep{kn.Ma98}. From the mother wavelet, $\psi _{0}$, we define a family of functions
\begin{equation}
\psi _{t,a}(s)=\frac{1}{\sqrt{a}}\psi _{0}\left( \frac{s-t}{a}\right).
\hspace{1cm}s>0,b\in \mathbb{R}
\end{equation}

Then for a given $\psi_{t,a}$, the wavelet scalogram, $A_{SCALO}$, has the affine kernel

\beq
\label{eq.wavelet_kernel}
\Psi_{SCALO}(\tau,\nu)=W_{\psi_{t,a}}(\tau,\nu) \eeq

where $W_{\psi_{t,a}}(\tau,\nu)$ is the WVD of $\psi_{t,a}$.

The choice of mother wavelet depends on the signal to be
analysed and the type of analysis desired. If the shape of the
signal to be detected is known, then the wavelet with the nearest
matching shape should be chosen \citep{kn.Me93}. Considering the
often (amplitude modulated) sinusoidal shape of DNOs and QPOs,
smoothly oscillating wavelets such as the Morlet, Mexican Hat and
Paul wavelet are indicated, rather than the abrupt step function of
the Haar wavelet, or the spiky Daubechies wavelet.

An additional consideration is that in the scalogram of a complex wavelet, such as
the Morlet, phase information from the data is lost. A scalogram using a complex wavelet has constant power across the time duration of an oscillation, and the time locations of maxima and minima in the oscillation cannot be detected. For real wavelets, however, such as the Mexican Hat wavelet, phase and amplitude information is superposed in the
time-frequency plane and extrema are easily detected \citep{kn.Pe00} - a feature that is not present in any of the TFRs studied so far in this paper.

In order to explore fully the behaviour of the wavelet scalogram we thus investigate two smooth wavelets with very different analytical capabilities: the complex Morlet wavelet and the real Mexican Hat wavelet.

\subsubsection{The Morlet Wavelet}

The Morlet wavelet is described by

\begin{equation}
\psi(\eta)=\pi^{-\frac{1}{4}}e^{-\frac{1}{2}\eta^2}e^{(i\omega_0
\eta)}
\end{equation}

where $\eta=t/s$ is a non-dimensional time parameter and $\omega_0$
is the wavenumber \citep{kn.To98}. The Morlet wavelet is essentially a complex
exponential of frequency $\frac{\omega_0}{2\pi}$, modulated in
amplitude by a Gaussian window \citep{kn.Pe00}. We have chosen to use $\omega_0=6$ in our analysis, which gives a
Morlet wavelet which  has 5 oscillations. This is the most commonly
used value of $\omega_0$, but is also appropriate given the QPO
detection criteria which will be discussed in section \ref{s.VWHyi}.

Since the wavelet is localized in time and scale, the wavelet scalogram at a particular time $t'$
is affected only by signal values close to $t'$, in a radius that
depends on the width of the wavelet as determined by the scale $a$.
The cone of influence (COI) of a point $(t',s)$ on the time-scale plane is the set of all points
$(t'',s)$ for which $x(t')$ affects $A_{SCALO}(t,a)$. Finite data length means that for points near the beginning or end
of the data set, the support of $A_{SCALO}(t,a)$ effectively
includes a number of points with zero amplitude (i.e. points before
or after measurement started), and therefore has significantly
reduced power. \citet{kn.To98} define the $e$-folding time $\tau_s$ for the
wavelet as the time-radius at which the wavelet power first drops by
a factor of $e^{-2}$ at an edge. The time extent to which these `edge effects' affect the wavelet scalogram depends on the
scale of the wavelet and the type of wavelet; at shorter scales
(shorter periods) the effect is minimal, while at very long scales,
the effect is marked. For the Morlet wavelet, $\tau_s=\sqrt{2}s$ \citep{kn.To98}.

It is preferable to plot the wavelet scalogram as a function
of frequency rather than scale. For the Morlet wavelet, the equivalent Fourier period for a given scale $s$ is

\begin{equation}
\textrm{Period} \ =\frac{4\pi s}{\omega_0 + \sqrt{2+{\omega_0}^2}}
\end{equation}

\begin{figure}
  \includegraphics[width=3in]{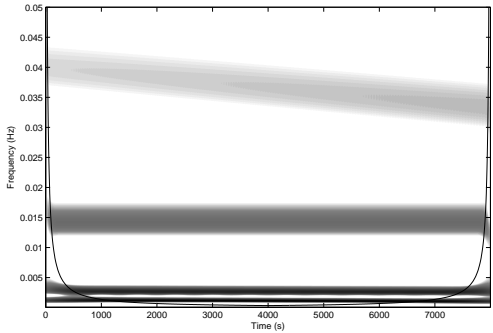}\\
  \caption{Morlet scalogram of the synthetic DN lightcurve.}\label{fig.Morlet_DN.eps}
\end{figure}

\citep{kn.Me93,kn.To98}.  Figure \ref{fig.Morlet_DN.eps} shows the Morlet scalogram of
the synthetic DN lightcurve. While all components are clearly visible, and there are no interference terms, the lpDNO and especially the DNO are not well resolved in frequency. The large black `U' marks the e-folding time at each scale: points outside this region are distorted by edge effects. It is clear that while edge effects are minimal at high frequencies, they become significant at low frequencies.

\begin{figure}
  \includegraphics[width=3in]{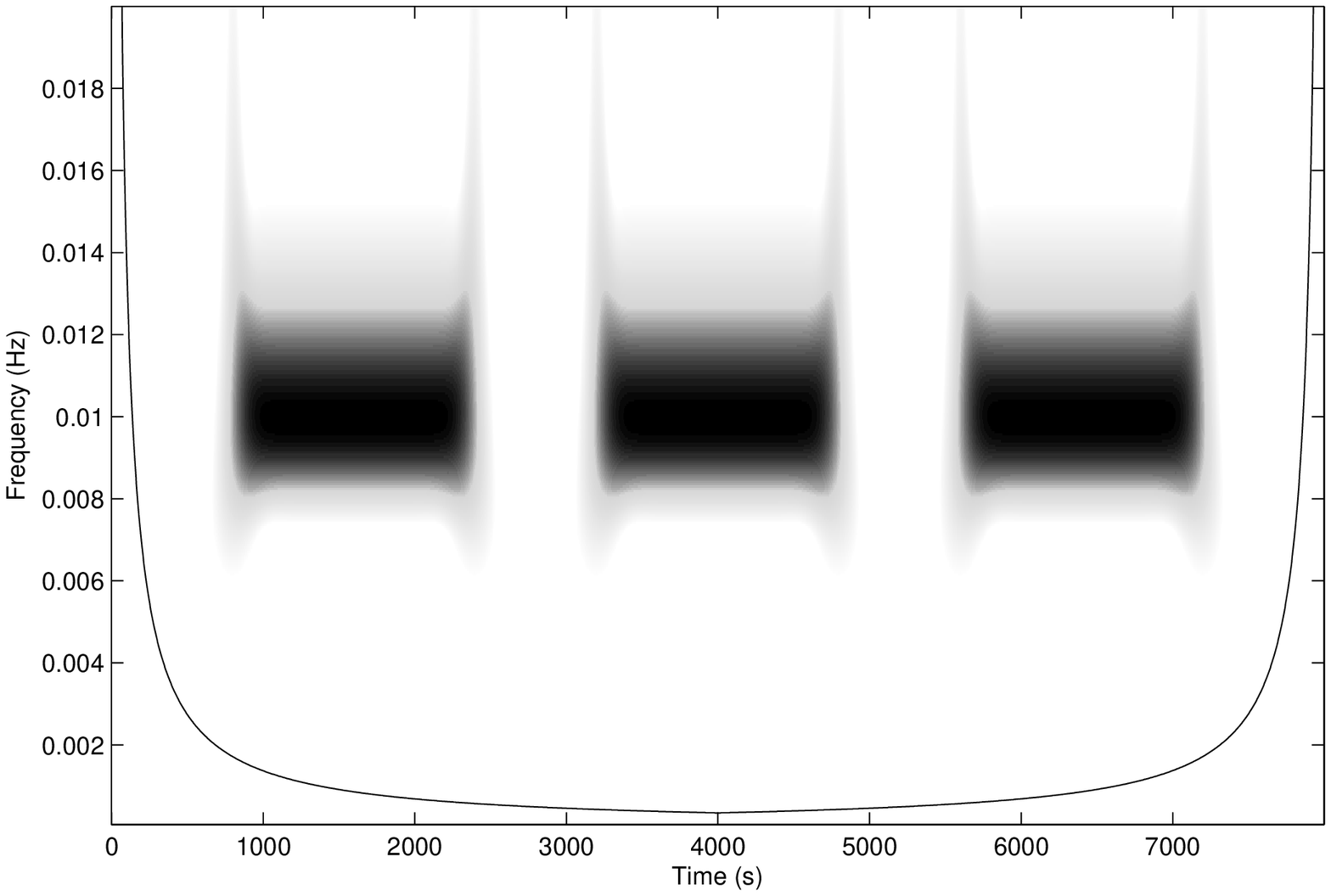}\\
  \caption{Morlet scalogram of the \un[100]{s} intermittent sinusoid.}\label{fig.Morlet_100.eps}
\end{figure}

\begin{figure}
  \includegraphics[width=3in]{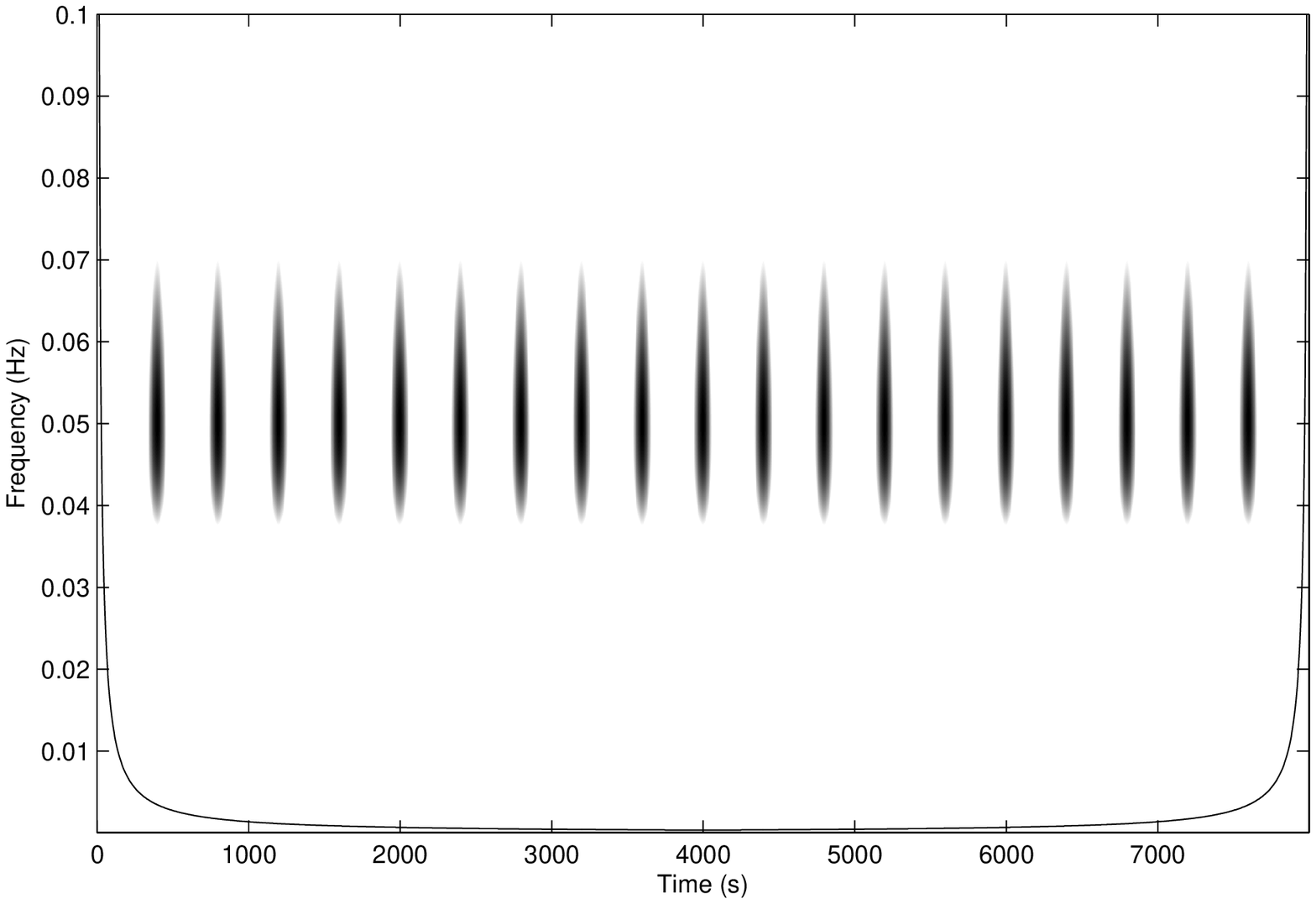}\\
  \caption{Morlet scalogram of the \un[20]{s} intermittent, amplitude modulated sinusoid.}\label{fig.Morlet_20.eps}
\end{figure}

In both the Morlet scalogram of the intermittent sinusoid with period \un[100]{s} (figure \ref{fig.Morlet_100.eps}) and the intermittent, amplitude-modulated sinusoid with period \un[20]{s} (figure \ref{fig.Morlet_20.eps}), we have an IT-free TFR in which the signal is well resolved in time, but this has come at the expense of poor frequency resolution: the Morlet scalogram has by far the worst frequency resolution of any of the TFRs investigated so far. There is, however, a way we can compensate for this poor frequency resolution, while maintaining the excellent time resolution and IT-free properties of this TFR.

\citet{kn.De92} show that for complex wavelets such as the Morlet,
the wavelet scalogram of a monocomponent signal with slowly
changing low frequency reaches a maximum at the instantaneous
frequency $\phi ^{\prime }(t)$ of the signal. \citet{kn.Ma98} extends this proof to
include the accurate measurement of the
instantaneous frequency of multiple components changing rapidly at
high frequency. Thus the points $(\bar{t},\bar{f})$ at which the Morlet scalogram $A_{MORLET}(t,f)$ has a local maximum in frequency, i.e. for which
\begin{equation}
\begin{array}{ccc}
\frac{\partial A_{MORLET}(t,f)}{\partial f}=0 & \textrm{and} & \frac{\partial ^2 A_{MORLET}(t,f)}{\partial f^2}<0, \\
\end{array}
\end{equation}

can be used to plot the instantaneous frequency of our signal accurately in the form of \emph{wavelet ridges}\citep{kn.Ma98}.

\begin{figure}
  \includegraphics[width=3in]{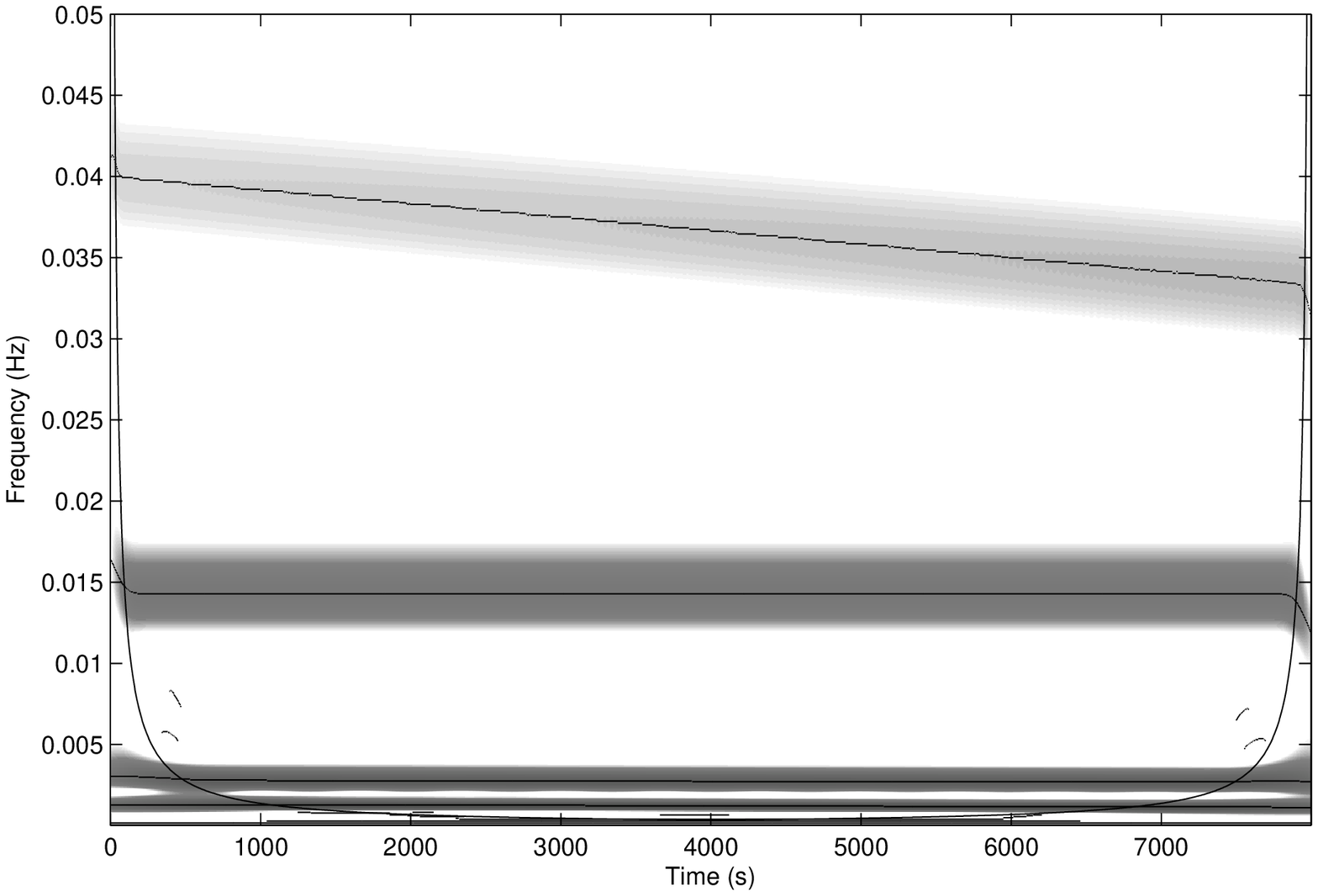}\\
  \caption{Morlet scalogram of the synthetic DN lightcurve, with ridges.}\label{fig.Morlet_DN_ridges.eps}
\end{figure}

\begin{figure}
  \includegraphics[width=3in]{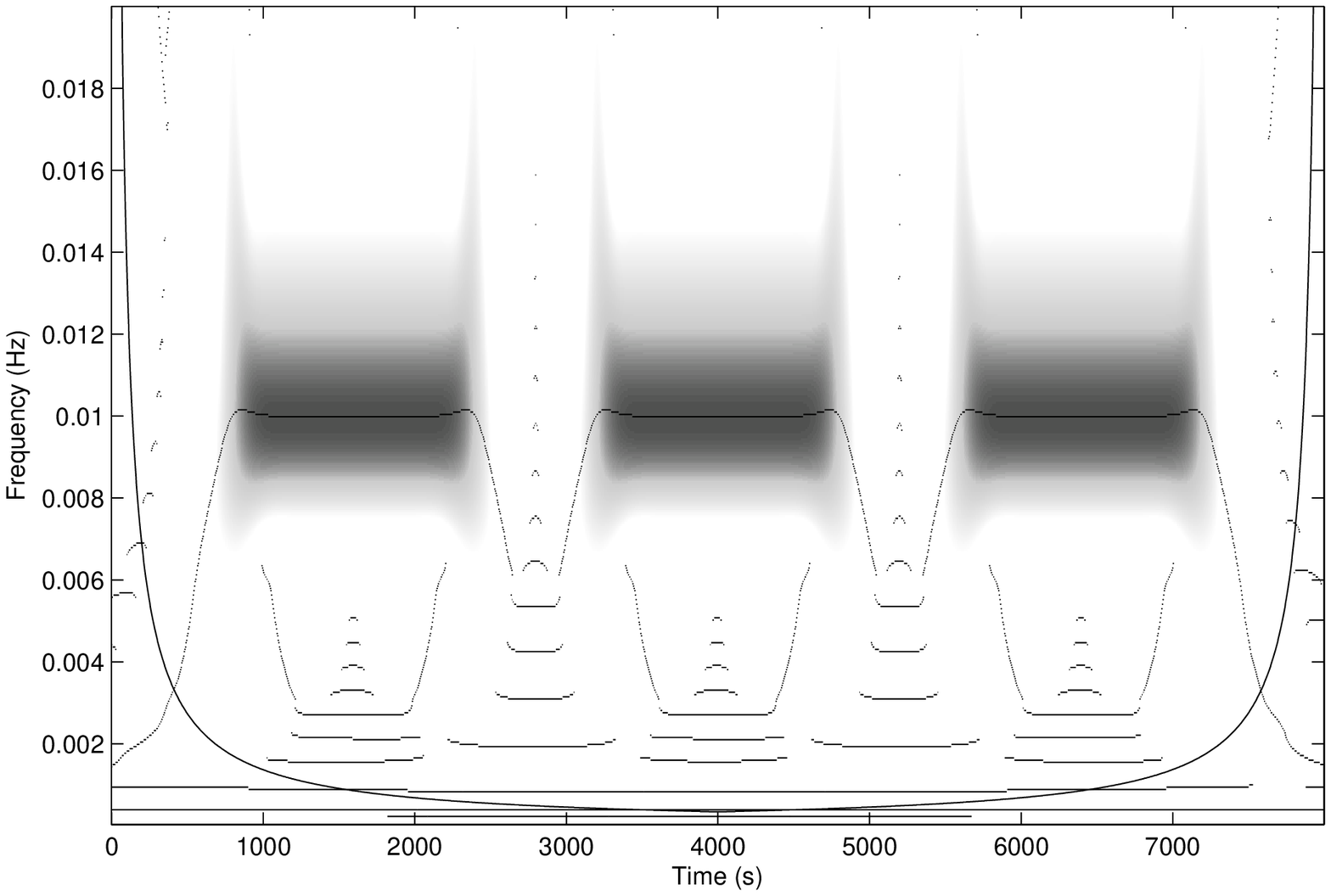}\\
  \caption{Morlet scalogram of the \un[100]{s} intermittent sinusoid, with ridges.}\label{fig.Morlet_100_ridges.eps}
\end{figure}

\begin{figure}
  \includegraphics[width=3in]{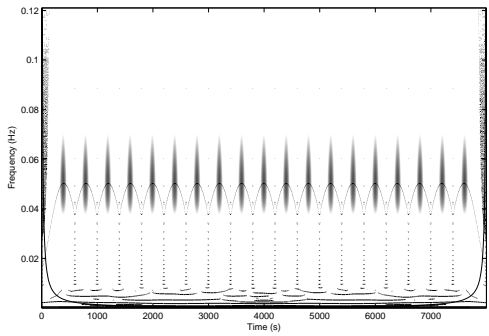}\\
  \caption{Morlet scalogram of the \un[20]{s} intermittent, amplitude modulated sinusoid, with ridges.}\label{fig.Morlet_20_ridges.eps}
\end{figure}

Figures \ref{fig.Morlet_DN_ridges.eps}, \ref{fig.Morlet_100_ridges.eps} and \ref{fig.Morlet_20_ridges.eps} again show the Morlet scalograms of our three test signals, but with the inclusion of wavelet ridges. Figure \ref{fig.Morlet_DN.eps} looks pleasingly close to the ideal TFR shown in figure \ref{fig.ideal_tfr}. Both figures \ref{fig.Morlet_100_ridges.eps} and \ref{fig.Morlet_20_ridges.eps} show wavelet ridges at frequencies other than the instantaneous frequency of the signal, but as these occur only at very low amplitudes they do not present the interpretational challenge that interference terms did in previous TFRs.

\subsubsection{The Mexican Hat Wavelet}

\begin{figure}
  \includegraphics[width=6in]{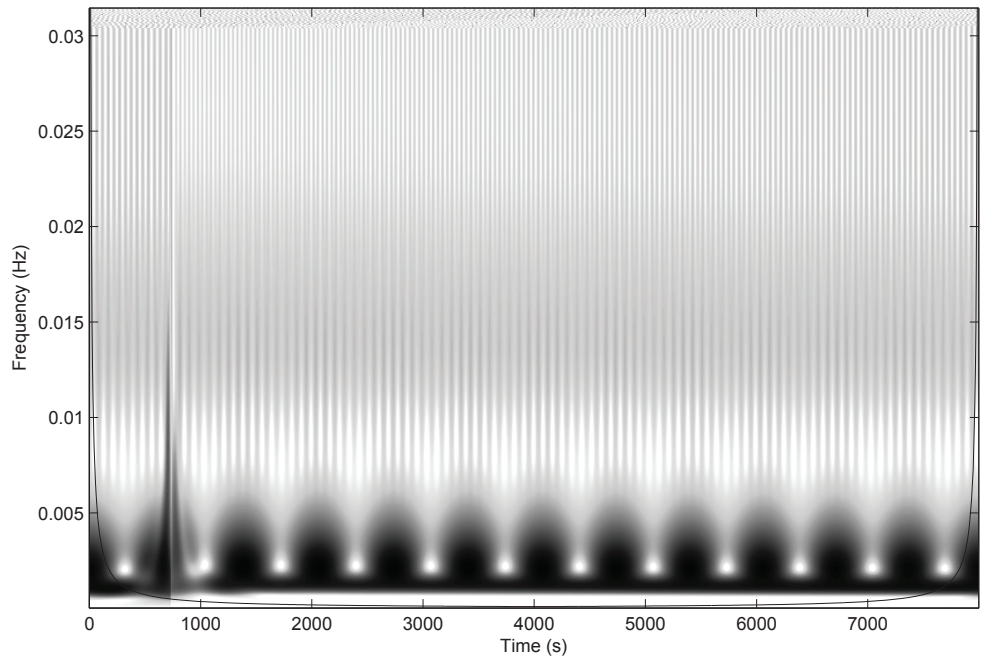}\\
  \caption{MHAT scalogram of the synthetic DN lightcurve.}\label{fig.MHAT_DN.eps}
\end{figure}

One of the most commonly used real wavelets is the Mexican Hat
Wavelet (MHAT), given by the normalized second derivative of a Gaussian
function:

\begin{equation}
\psi(\eta)=\frac{1}{\sqrt{\Gamma(2.5)}}(1-\frac{1}{\eta^2})e^{\frac{-\eta^2}{2}}
\end{equation}

where again $\eta=t/s$. The equivalent Fourier period for a given
scale $s$ is

\begin{equation}
\textrm{Period} \ =\frac{2\pi s}{\sqrt{2.5}}.
\end{equation}

which effectively increases the minimum observed time-scale from $2dt$ to $7.95dt$ \citep{kn.Mc07}. As a result, the MHAT wavelet is not useful for examining very short period phenomena. Figure \ref{fig.MHAT_DN.eps} shows the MHAT scalogram of the synthetic DN signal. Notice that the highest frequency achievable is \un[0.03]{Hz}, which is below the starting frequency of the DNO. It is clear that the frequency resolution of the MHAT wavelet is considerably worse even than that of the Morlet wavelet. The usefulness of the MHAT wavelet is, however, apparent in the appearance of the lpDNO and the QPOs in the scalogram: individual extrema are discernable (the wavelet scalogram is an entirely positive TFR, so all signal extrema appear as maxima in the wavelet scalogram \citep{kn.Ma98}).

\begin{figure}
  \includegraphics[width=6in]{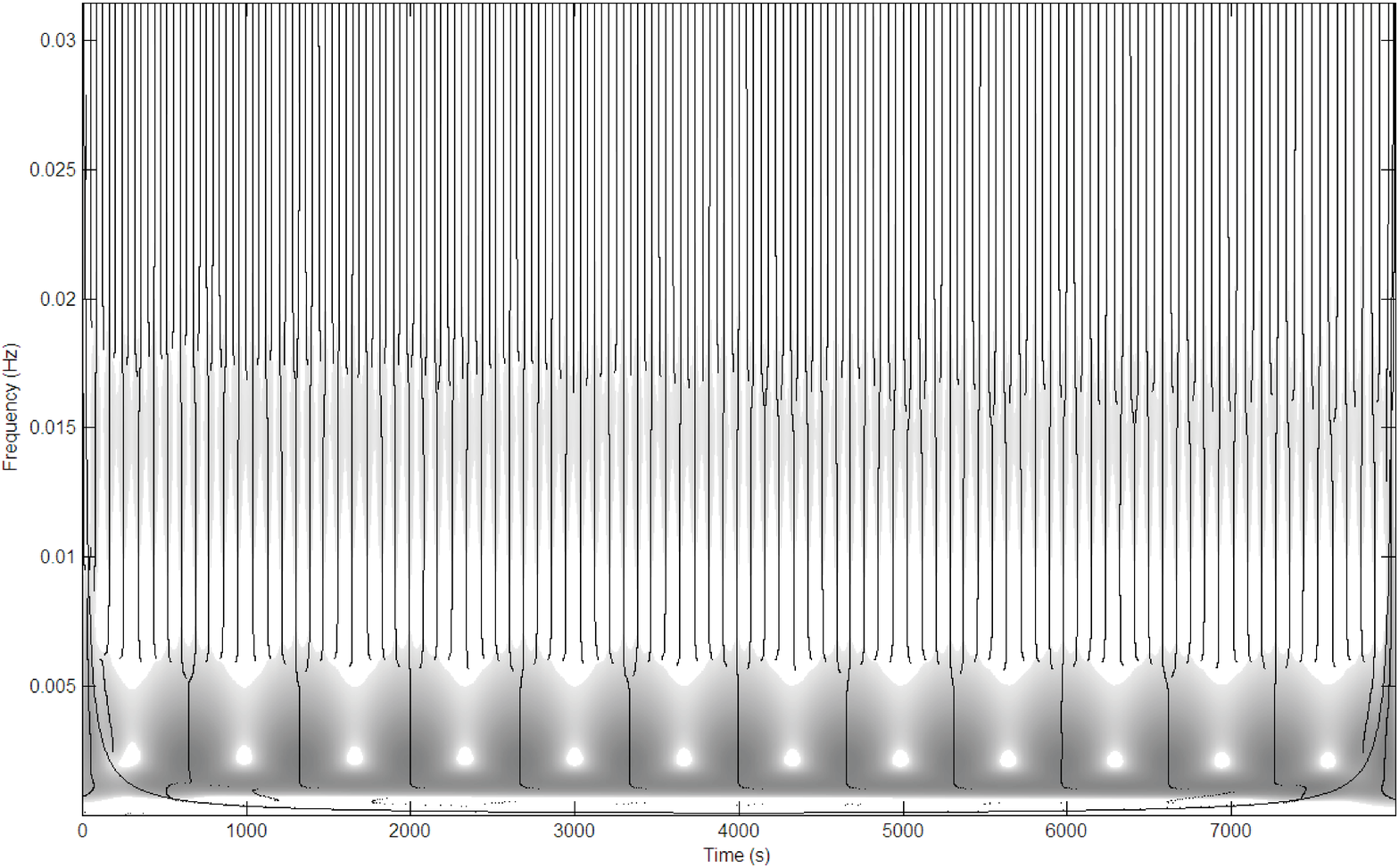}\\
  \caption{MHAT scalogram of the synthetic DN lightcurve with ridges.}\label{fig.MHAT_DN_ridges.eps}
\end{figure}

\begin{figure}
  \includegraphics[width=6in]{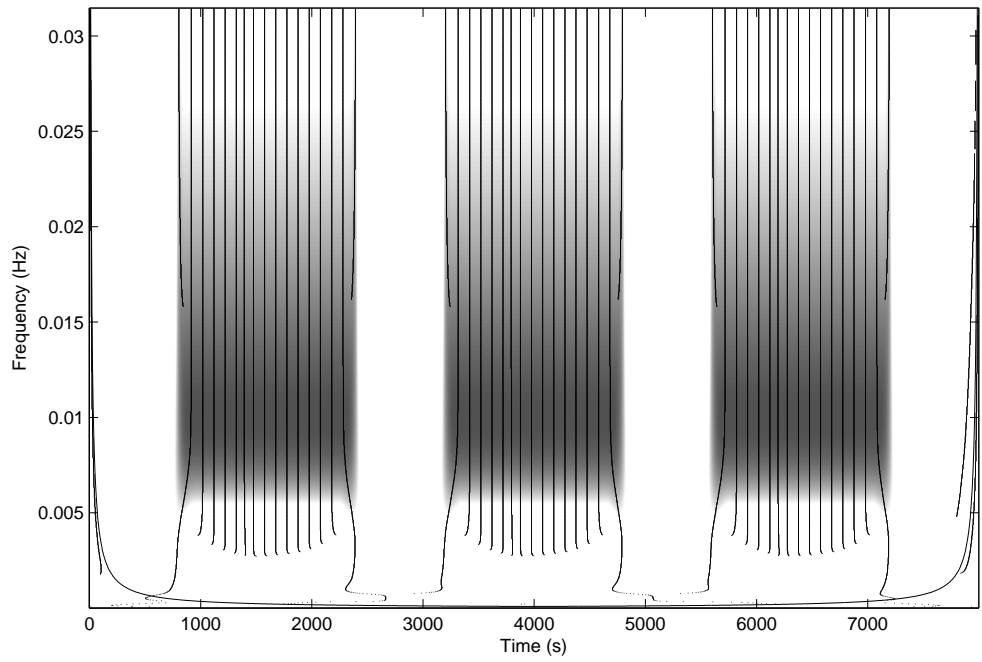}\\
  \caption{MHAT scalogram of the 100s intermittent sinusoid with ridges.}\label{fig.MHAT_100_ridges.eps}
\end{figure}

Similar to the wavelet ridges of the Morlet wavelet, the points $(\bar{t},\bar{f})$ at which the MHAT scalogram $A_{MHAT}(t,f)$ has a local maximum in time, i.e. for which
\begin{equation}
\begin{array}{ccc}
\frac{\partial A_{MHAT}(t,f)}{\partial t}=0 & \textrm{and} & \frac{\partial ^2 A_{MHAT}(t,f)}{\partial t^2}<0, \\
\end{array}
\end{equation}
form \emph{maxima lines} which identify maxima across frequencies \citep{kn.Ma98}. Maxima lines are traditionally used to detect singularities and discontinuities. \citet{kn.Po03} and \citet{kn.Mc07}, for example, use maxima lines to investigate the sunspot index and solar flare x-ray emission respectively. For our data, however, maximal lines can be used to detect the extrema in quasi-periodicities, which can be used to measure coherence, investigate phase relationships between different frequency phenomena and construct O-C diagrams. Figures \ref{fig.MHAT_DN_ridges.eps} and \ref{fig.MHAT_100_ridges.eps} show the MHAT scalogram of the synthetic DN light curve and the \un[100]{s} intermittent sinusoid respectively, with the addition of maxima lines.

\begin{figure}
  \includegraphics[width=6in]{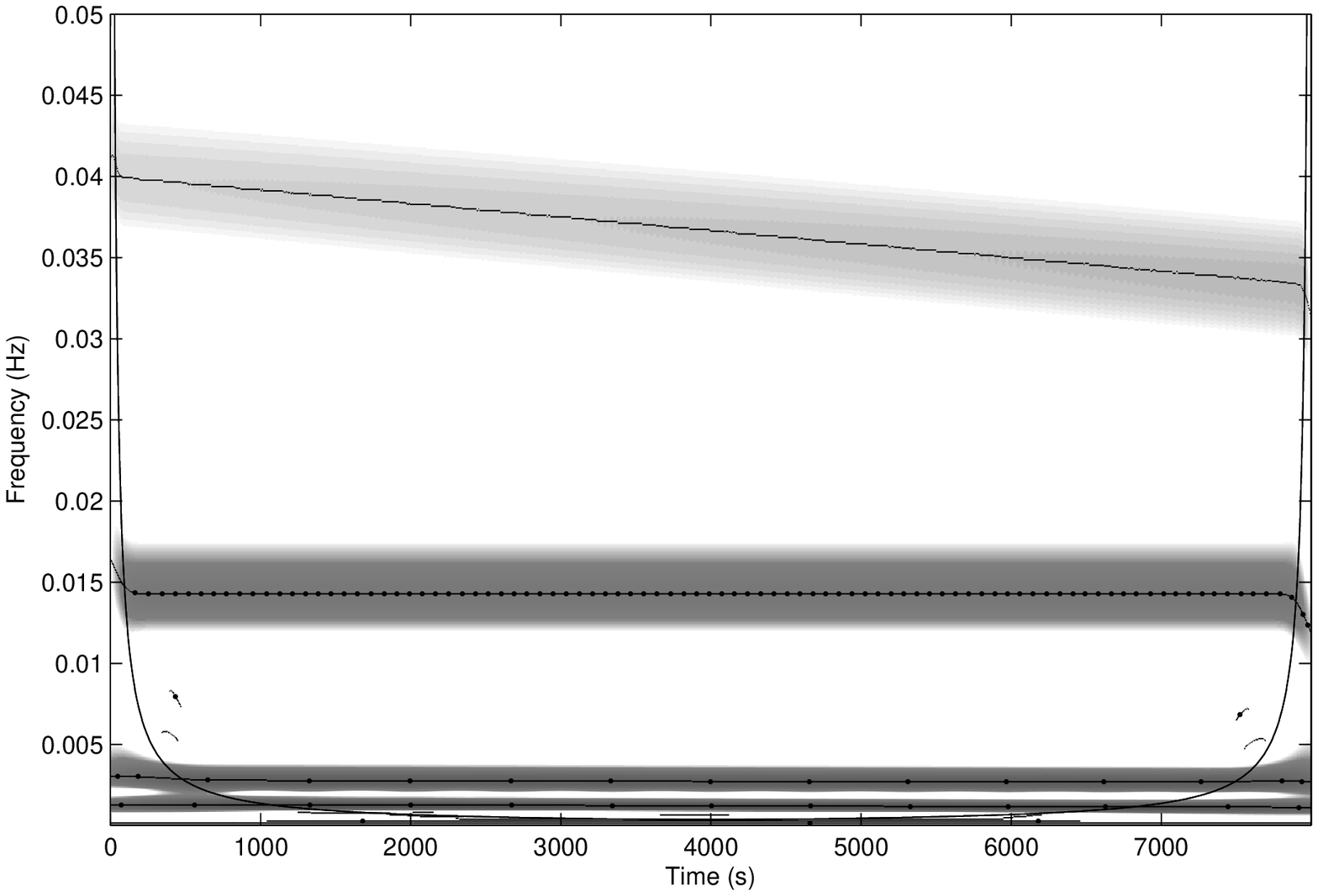}\\
  \caption{Morlet ridged scalogram of the synthetic DN lightcurve with wavelet ridges and maxima points.}\label{fig.Both_DN_ridges.eps}
\end{figure}

Both the Morlet scalogram with the addition of wavelet ridges and the MHAT scalogram with the addition of maxima lines provide valuable information in the analysis of our data. To facilitate easy analysis, we combine the information provided by the wavelet ridges and maxima lines on a single `Morlet ridged scalogram': we plot all wavelet ridges, but only those points of the maxima lines which intersect with wavelet ridges. Figure \ref{fig.Both_DN_ridges.eps} shows the our final TFR of the synthetic DN signal. Here we finally have a TFR in which all frequency components can be seen clearly, as well as the times at which each component has an extremum.

At least in the case of noise-free signals, it is clear that the Morlet ridged scalogram is the TFR best suited to our needs, providing excellent frequency resolution over a broad frequency ranges, combined with the time resolution needed to resolve intermittent signals. However, when analysing real data the statistical properties of the TFR are important for judging the likelihood that a detected oscillation is a true signal, and not an artefact of the noise. It thus behoves us to investigate the statistical properties of each of our six kernels, before declaring a winner.

\subsection{Statistical Properties of TFRs}
\label{ss.statistics}

We assume that we have a non-stationary
deterministic signal superimposed on weakly stationary\footnote{A
time series is \emph{weakly stationary} if all realizations
$\{x_t\}$ of the process $\{X_t\}$ have the same mean, the
(auto)covariance between any 2 samples from the same realization is
a function of the time-lag between the samples only, and the
variance is finite \citep{kn.Ke83}.} non-white noise. This is not an unreasonable assumption: the noise spectrum of our data changes very slowly if at all over the duration of an observing run. We will
discuss methods of estimating the power spectrum of the noise in section \ref{ss.power_spectrum_estimation}, but
note here that the assumption of weak stationarity ensures that the
periodogram can be used as an estimator for the power spectrum of
the noise \citep{kn.Ti95}.

The WVD of a realization of a stochastic signal is an estimator for
the theoretical WVD of the generating process \citep{kn.Me97}. For a
weakly stationary noise process with power spectrum $S(f)$, the
expected value of the WVD, $E[W_x(t,f)]$, is given by the power
spectrum of the process:
\begin{equation}
 E[W_x(t,f)]=S(f)
\end{equation}
\citep{kn.Qi93}. For a deterministic signal corrupted with
independent, identically distributed (i.i.d) gaussian noise with
power spectrum $S_{noise}(f)$, the expected value of the WVD is the
sum of the WVD of the signal and the spectrum of the noise:
\begin{equation}
E[W_x(t,f)]=W_{signal}+ S_{noise}(f)
\end{equation}
\citep{kn.Fl84}. The WVD is thus a biased estimator of the signal,
and the larger $S_{noise}(f)$, the worse the bias. The WVD of the
noisy signal, since it is determined by the power spectrum of the
noise, is also not consistent: $ W_x(t,f) \sim
S_{noise}(f)\frac{\chi_2^2}{2}$.

Thus the WVD suffers from the same estimation problems as the
periodogram. However, by estimating the power spectrum of the noise,
confidence contours (the 2-D equivalent of confidence levels) can be
used to determine regions in the time-frequency plane at which the
signal has power significantly above that expected for pure noise
(see \citet{kn.To98} for a detailed discussion of confidence
contours for the wavelet scalogram).

The expected values of any Cohen TFR of a deterministic signal
corrupted with i.i.d gaussian noise are also given by the sum of the
noise-free TFR and the power spectrum of the noise, as was the case
for the WVD. Thus these TFR estimators, which include the CWD, the
BJD and the Gabor spectrogram, are also biased, with the amount of
the bias depending on the noise level. We also have $C_x(t,f) \sim
S_{noise}(f)\frac{\chi_2^2}{2}$ \citep{kn.Po90}. In general, the
variance of the TFR will depend on the amount of smoothing by the
kernel: more smoothing results in a lower variance \citep{kn.Wi98}.
The BJD is the minimum variance distribution for white noise
\citep{kn.He95}, while \citet{kn.Wi98} shows that the Gabor spectrogram,
when a rectangular window is used, gives the best performance of all
Cohen TFRs in the presence of non-white noise, as it has the
smallest variance.

For the SPWVD with $L_g=M$, for a deterministic signal corrupted by
i.i.d gaussian noise, $C_{SPWVD}(t,f) \sim
\frac{2}{2M-1}S_{noise}(f)\frac{\chi_2^2}{2}$ \citep{kn.Fl84}. Thus
the variance of the SPWVD decreases as the length of the
time-smoothing window increases.

For the ASPWVD, with $L_g$ at scale
$s$ given by $M_s$, then for a deterministic signal corrupted by
i.i.d gaussian noise,

\begin{equation}
A_{ASPWVD}(t,s)\sim \frac{2}{2M_s-1}S_{noise}(s)\frac{\chi_2^2}{2}
\end{equation}

\citep{kn.Fl84}. Thus the variance of the ASPWVD decreases as the
window length increases, as in the case of the SPWVD, but also
depends on the analysing scale $s$.

Since the CKR's integrals with respect to frequency, and over the
entire time-frequency plane, are zero (because
$\Psi_{CKR}(\tau,\nu)$ is zero on the entire $\nu$-axis), the CKR is
 not an energy distribution in the statistical sense (which is
why we refer to it as a representation, rather than a distribution)
\citep{kn.Hl95}. This means that its expected value is not related to the power spectrum as for the other TFRs. The CKR of a deterministic signal corrupted by
i.i.d gaussian noise is the sum of the CKR of the noise-free signal
and $D(t,f;g)$, where

\beq D(t,f;g)=\int_{-\infty}^{\infty}g(\tau)|\tau|r(\tau)e^{-i2\pi f
\tau}d\tau \eeq

and $r(\tau)$ is the auto-covariance of the noise. This means that
for white noise, $D(t,f;g)=0$, and the CKR is an unbiased estimator
\citep{kn.Oh92}. However, \citet{kn.He95} show that of all the Cohen
TFRs the CKR has the highest variance, and hence the least desirable
performance in the presence of noise. In addition, confidence
contours cannot easily be constructed for the CKR since the
stochastic distribution of the CKR is not calculable from the power
spectrum as is the case for the other TFRs.

\citet{kn.Pe95} shows that the time integral of the wavelet
scalogram can be used as an unbiased, consistent estimator for the
power spectrum. It is also the minimum affine variance estimator,
and we have $A_{SCALO}\sim \frac{1}{2}S_{noise}\chi_2^2$. Thus from a statistical point of view, the wavelet scalogram is also the best TFR to use for our data.

\clearpage

\section{Application of Time-Frequency Techniques to VW Hyi}
\label{s.VWHyi}

In this section we use the Morlet ridged scalogram introduced in the
previous section to analyse quiescent and outburst lightcurves from VW Hyi.
Our purpose is threefold. Firstly, since all of these runs have been
analysed in detail previously, we see how previously detected DNOs
and QPOs appear in the wavelet scalogram, and hence build up a set
of criteria for objectively detecting QPOs using the wavelet
scalogram. We then use these criteria to see if there are QPOs in
the VW Hyi data which have not been identified by the old methods.
Finally, we use the wavelet methods to investigate the non-stationary behaviour of all
DNOs and QPOs detected.

This analysis establishes the reliability of the Morlet ridged scalogram (we find that all previously identified DNOs and QPOs are detected by the scalogram), and provides an opportunity
to compare the information elicited by the new techniques with that
given by traditional techniques such as the Fourier Transform and
O-C curves.

\subsection{Observations}

VW Hydri is an SU UMa dwarf nova with an outburst time scale of the
order of 28 days \citep{kn.Wa95}. Since it is circumpolar at the
Sutherland site of the South African Astronomical Observatory, and
is relatively bright (9-\un[13]{mag}), VW Hyi presents an extended
season for maximum time-resolution observation. In addition to
quiescent observations, many observations have been made selectively
towards the end of outburst as the star returns to quiescence
specifically to study DNO and QPO behaviour \citep{kn.Wa06a}, and
several groups of runs taken on consecutive nights allow for the
study of persistent or evolutionary phenomena across nights.

Our VW Hyi data are from the UCT archive, and have been previously
analysed in \citet{kn.Wo02a}, \citet{kn.Wo02b}, \citet{kn.Wa03},
\citet{kn.Wa06a} \citet{kn.Pr06} and \citet{kn.Wa08}. Details of the
individual observing runs for outburst and quiescent observations
are given in tables \ref{tbl:VWHyi_outburst_runs} and
\ref{tbl:VWHyi_quiescent_runs} respectively. All observations were
taken at the Sutherland site of the South African Astronomical
Observatory, using the 20'', 40'' or 74'' telescope as indicated.
Data acquisition for observations on the 20'' and 40'' telescope up
to and including run s5248 used the UCT photometer \citep{kn.Wa71a};
subsequent runs on this telescope and the 74'' were made using the
University of Cape Town CCD \citep{kn.Do95} operating in
frame-transfer mode. Most CCD observations were taken in white
light; observations made with a B filter are marked with an
asterisk.

The data were pre-whitened by removing the mean and first-order trend. VW Hyi lightcurves also frequently show a strong orbital hump, which can dominate the periodogram and wavelet scalogram. To avoid this, and enable smaller amplitude periodicities to be examined, we removed a sinusoid and first harmonic at the orbital hump frequency.

\subsection{Power Spectrum Estimation}
\label{ss.power_spectrum_estimation}

\citet{kn.To98}'s statistical detection of significant oscillations in the wavelet scalogram requires estimation of the periodogram of the noise spectrum of the signal, a perennial problem in noisy astronomical time series analysis. \citet{kn.Va05} suggests fitting a linear function to the log-periodogram, since his data show a red-noise power spectrum well modeled by power law
(i.e. linear in log-space). Flickering also gives our data a red noise spectrum, but
CV flickering has less power at very low frequencies than the X-ray
data of \citet{kn.Va05}; we therefore fit a low-order polynomial
(generally between third and fifth-order) in log-space, using an OLS
fitting programme.

To check that our assumption of a stationary noise structure for
each data set was correct, we calculated the best-fit noise-model
for sections of each run as well as for the entire run; in all cases
we found that the models for each section were very similar to the
model for the complete time series.

\subsection{Analysis}
\begin{figure}
  \includegraphics[width=5in]{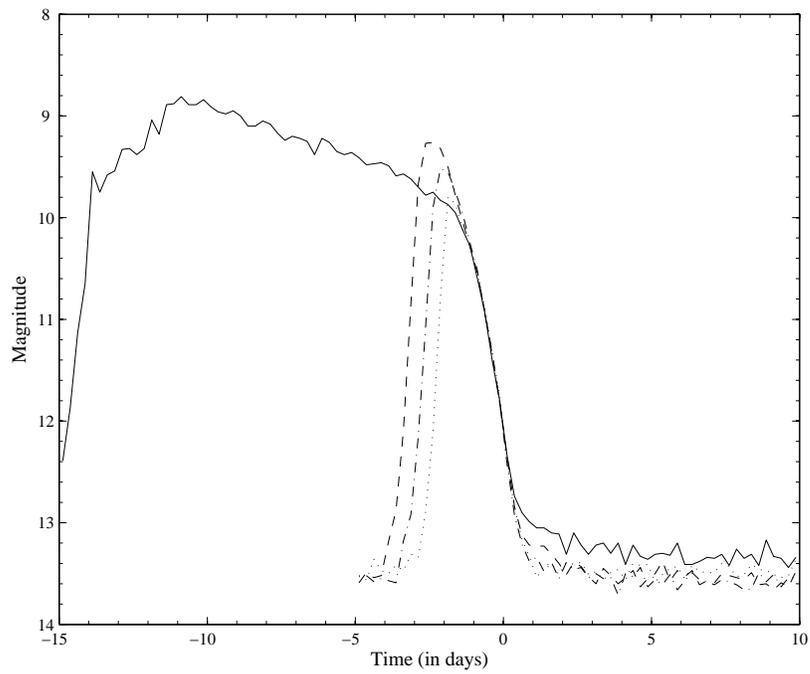}\\
  \caption{Averaged outburst profiles for
the four different types of outburst seen in VW Hyi. Solid line
indicates superoutburst, dashed line indicates long normal outburst,
dash-dot indicates medium normal and dotted indicates short normal
outburst.}\label{fig:outburst_profile}
\end{figure}

VW Hyi outbursts can be divided into four groups: superoutburst,
long normal outburst, medium normal outbursts and short normal
outbursts \citep{kn.Wo02a}. The RASNZ VW Hyi data, spanning
1972-2004, have been used to construct average outburst profiles
(figure \ref{fig:outburst_profile}), onto which each outburst
observation can be placed \citep{kn.Wo02a}. Each observation is thus
assigned a phase (column `Start' in tables
\ref{tbl:VWHyi_outburst_runs} and \ref{tbl:VWHyi_quiescent_runs}),
relative to the arbitrary $T=$\un[0]{d} (column `T=0' in tables
\ref{tbl:VWHyi_outburst_runs} and \ref{tbl:VWHyi_quiescent_runs}) of
the template. We have assigned quiescent observations a phase
relative to $T=$\un[0]{d} of the next outburst, as we have several
quiescent observations occurring a few days before an outburst for
which we also have data.

We computed the Morlet ridged scalogram for each run to see if previously detected QPOs were detected by the scalogram. We found that all previously detected QPOs were visible in the Morlet
ridged scalogram, with power above the 95\% confidence level and a single
continuous instantaneous frequency line of stable or slowly changing frequency. We
counted the number of extrema (given by the maximal points) which
occurred during the time span of the confidence contour, and found
that while 3 cycles (6 maximal points) were sometimes sufficient to
have been identified as a QPO in the periodogram (e.g. run s0030),
usually 5 or more cycles were present. We thus defined a QPO as at
least 5 cycles (10 maximal points in the MHAT) at power above the
95\% confidence level (on the Morlet spectrum), or 6 significant
cycles spread over two confidence contours at similar frequency. We
only deviated from this definition when there was a clear triplet of
QPOs with fundamental, first and second harmonic, and one of these
contained less than 5 cycles, in which case it was identified as a
QPO (e.g. run s1307). We used the instantaneous frequency to find the (average) period of the detected QPO.

\begin{deluxetable}{lccccrrr}
\tabletypesize{\small}
\tablecaption{VW Hyi: Outburst observations from the UCT archive. Runs marked with an asterisk were taken using a B filter.\label{tbl:VWHyi_outburst_runs}}
\tablehead{\colhead{Run} & \colhead{Telescope} & \colhead{Length} & \colhead{Int} & \colhead{Outburst} & \colhead{HJD Start} & \colhead{$T=0$} & \colhead{$T$ Start}\\ & & \colhead{(hrs)}& \colhead{time (s)} &\colhead{Type} &\colhead{$+2440000$} &\colhead{(d)} & \colhead{(d)}\\}
\startdata
s0018 & 20 & 2.04 & 2 & L & 1572.52 & 1572.75 & -0.23 \\
s0019 & 20 & 4.30 & 5 & L & 1573.44 & 1572.75 & 0.69 \\
s0026 & 20 & 3.32 & 5 & L & 1574.50 & 1572.75 & 1.75 \\
s0030 & 20 & 6.92 & 5 & L & 1575.47 & 1572.75 & 2.72 \\
s0110 & 40 & 2.47 & 2 & Super & 1662.30 & 1677.20 & -14.90 \\
s0111 & 40 & 0.97 & 2 & Super & 1662.41 & 1677.20 & -14.79 \\
s0112 & 40 & 2.31 & 2 & Super & 1662.48 & 1677.20 & -14.72 \\
s0122 & 40 & 2.56 & 5 & Super & 1675.33 & 1677.20 & -1.87 \\
s0124 & 40 & 1.84 & 5 & Super & 1676.37 & 1677.20 & -0.83 \\
s0127 & 40 & 3.77 & 4 & Super & 1677.29 & 1677.20 & 0.09 \\
s0129 & 40 & 1.86 & 5 & S & 1691.30 & 1690.20 & 1.10 \\
s0480 & 40 & 1.73 & 5 & Super & 2017.38 & 2022.65 & -5.27 \\
s0484 & 40 & 3.91 & 4 & Super & 2023.31 & 2022.65 & 0.66 \\
s1277 & 40 & 1.78 & 4 & M & 2352.39 & 2354.20 & -1.81 \\
s1307 & 40 & 1.33 & 4 & M & 2354.39 & 2354.20 & 0.19 \\
s1322 & 40 & 4.91 & 4 & M & 2355.39 & 2354.20 & 1.19 \\
s1571 & 40 & 1.99 & 4 & Super & 2402.32 & 2403.40 & -1.08 \\
s1616 & 40 & 2.27 & 3 & Super & 2404.31 & 2403.40 & 0.91 \\
s2241 & 40 & 2.63 & 5 & Super & 2758.41 & 2770.00 & -11.59 \\
s2243 & 40 & 1.65 & 5 & Super & 2759.40 & 2770.00 & -10.60 \\
s2623 & 40 & 3.60 & 5 & M & 3515.30 & 3514.60 & 0.70 \\
s2911 & 40 & 3.14 & 4 & L & 4934.43 & 4937.10 & -2.67 \\
s2915 & 40 & 0.73 & 4 & L & 4937.28 & 4937.10 & 0.18 \\
s2917 & 40 & 2.18 & 4 & L & 4939.49 & 4937.10 & 2.39 \\
s3078 & 40 & 1.95 & 4 & Super & 5323.34 & 5328.50 & -5.16 \\
s3416 & 40 & 1.19 & 2 & M & 5967.55 & 5967.00 & 0.55 \\
s5248 & 40 & 4.66 & 5 & M & 8202.36 & 8201.75 & 0.61 \\
s6138\_I & 40 & 2.09 &  4 & M & 11898.28 & 11897.85 & 0.43 \\
s6138\_II & 40 & 3.90 & 5 & M & 11898.43 & 11897.85 & 0.58 \\
s6184 & 40 & 1.68 & 5 & L & 11957.28 & 11957.25 & 0.03 \\
s6316 & 40 & 5.72 & 5 & L & 12354.24 & 12353.95 & 0.29 \\
s6528\_I & 74 & 5.81 & 4 & L & 12520.22 & 12519.65 & 0.57 \\
s6528\_II & 74 & 4.18 & 5 & L & 12520.47 & 12519.65 & 0.82 \\
s7222\_I* & 74 & 3.61 & 4 & Super & 13007.31 & 13006.40 & 0.91 \\
s7222\_II* & 74 & 3.03 & 4 & Super & 13007.46 & 13006.40 & 1.06 \\
s7301 & 40 & 2.94 & 5 & M & 13087.24 & 13086.10 & 1.14 \\
s7311* & 74 & 5.41 & 4 & M & 13139.19 & 13138.95 & 0.24 \\
s7342 & 40 & 3.07 & 5 & M & 13155.19 & 13154.90 & 0.29 \\
s7621 & 40 & 4.03 & 6 & L & 13463.23 & 13463.10 & 0.13\\
\enddata
\end{deluxetable}

\begin{deluxetable}{cccccccc}
\tabletypesize{\small}
\tablecaption{VW Hyi: Quiescent observations from the UCT archive\label{tbl:VWHyi_quiescent_runs}}
\tablehead{\colhead{Run} & \colhead{Telescope} & \colhead{Length} & \colhead{Int} & \colhead{HJD Start} & \colhead{$T=0$} & \colhead{$T$ Start}& \colhead{HJD Next}\\ & & \colhead{(hrs)}& \colhead{time (s)}  &\colhead{$+2440000$} &\colhead{(d)} & \colhead{(d)}&\colhead{$+2440000$}\\}
\startdata
s0073 & 40 & 2.83 & 5 & 1648.25 & 1613.00 & -28.95 & 1677.20 \\
s0077 & 40 & 4.11 & 5 & 1602.44 & 1580.00 & -10.56 & 1613.00 \\
s0085 & 40 & 1.67 & 5 & 1604.44 & 1580.00 & -8.56 & 1613.00 \\
s0093 & 40 & 3.33 & 5 & 1605.45 & 1580.00 & -7.54 & 1613.00 \\
s0102 & 40 & 1.31 & 5 & 1657.33 & 1613.00 & -19.87 & 1677.20 \\
s0105 & 40 & 1.80 & 2 & 1660.29 & 1613.00 & -16.91 & 1677.20 \\
s1414 & 40 & 2.89 & 4 & 2384.30 & 2370.00 & -19.10 & 2403.40 \\
\enddata
\end{deluxetable}

We give a detailed discussion of several of the runs analysed, showing how DNOs and QPOs appear in real data and highlighting the additional information which the Morlet ridged scalogram makes available. Discussions of further runs are available in \citet{kn.Bl08}.

\subsubsection{s0110, s0122 and s0127}

\begin{figure}
  \includegraphics[width=6in]{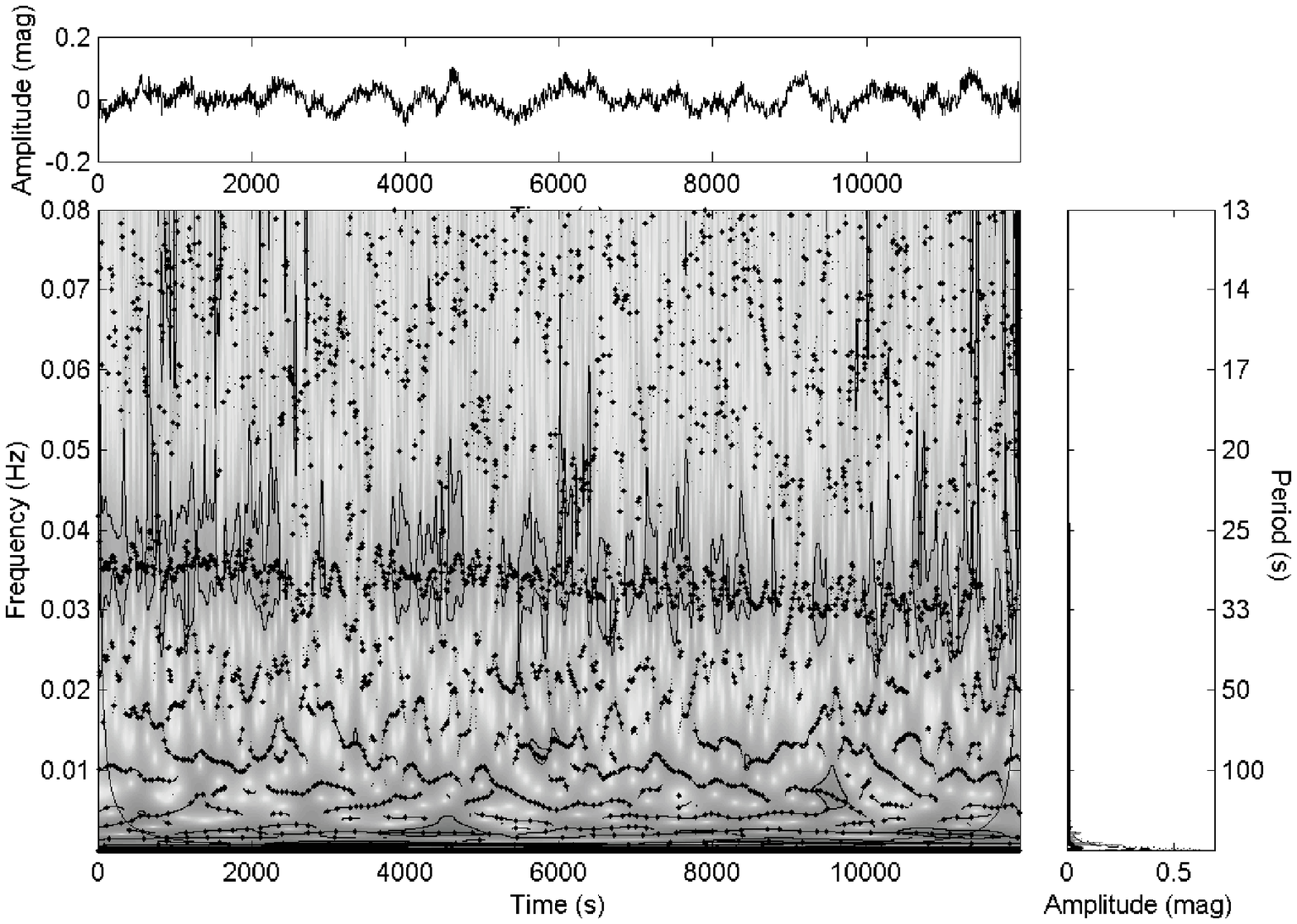}\\
  \caption{Prewhitened lightcurve, Morlet ridged scalogram and periodogram of run s0127.}\label{fig.s0127_0_08.eps}
\end{figure}

\begin{figure}
  \includegraphics[width=6in]{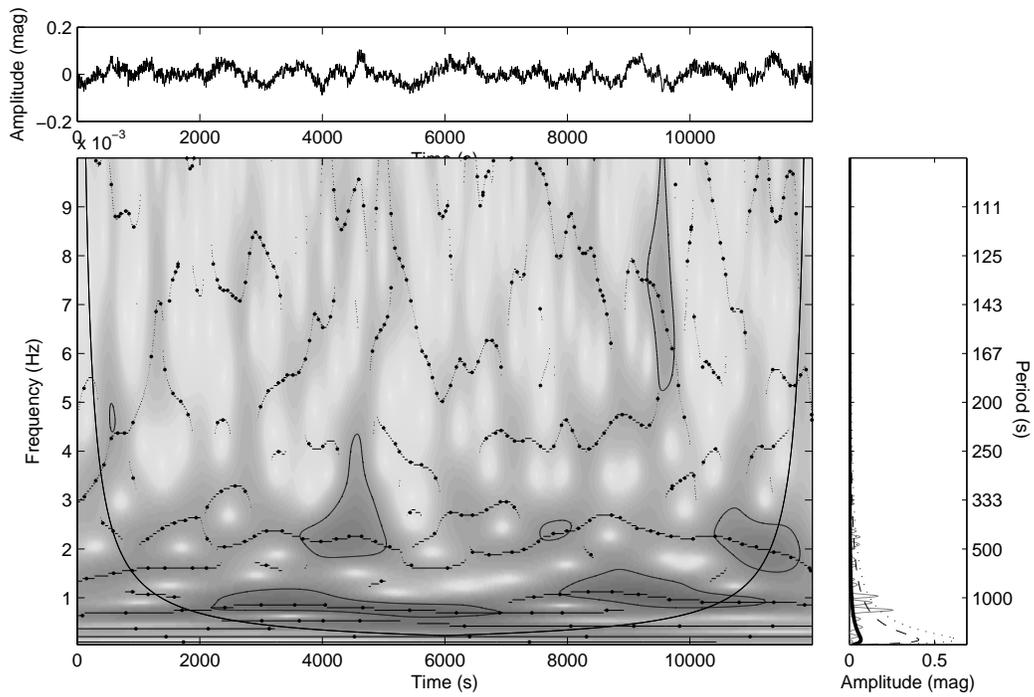}\\
  \caption{Prewhitened lightcurve, Morlet ridged scalogram and periodogram of run s0127 for the frequency range 0 - \un[0.01]{Hz}, showing QPOs.}\label{fig.s0127_0_01.eps}
\end{figure}

These three runs are part of a series of runs spanning a superoutburst. Run s0110 was taken on during the rise to supermaximum ($T=$\un[14.90]{d}), while runs s0122 and s0127 were taken on consecutive nights during decline, from $T=-0.83$ to \un[0.09]{d}.

s0127 (figures \ref{fig.s0127_0_01.eps} and \ref{fig.s0127_0_08.eps}) exhibits a plethora of phenomena, and is an excellent example of both the broad frequency range covered by DNOs and QPOs (from periods of \un[15]{s} to \un[1250]{s}) and the intermittent behaviour of these phenomena which have necessitated the development of the Morlet ridged scalogram. Figure \ref{fig.s0127_0_08.eps} clearly shows the \un[27]{s} DNOs, increasing in
period over the duration of the run to \un[34]{s}. There are also
suggestions of the \un[15]{s} first harmonic for the first
\un[3000]{s}, and again near the end of the run. Figure \ref{fig.s0127_0_01.eps} shows the \un[450]{s} QPO
found by \citet{kn.Wa06a}, as well as evidence of a
\un[1250]{s} oscillation, changing to \un[1075]{s} after 8000
seconds, which we categorize as a subharmonic of the \un[480]{s}
oscillation.

\begin{figure}
  \includegraphics[width=6in]{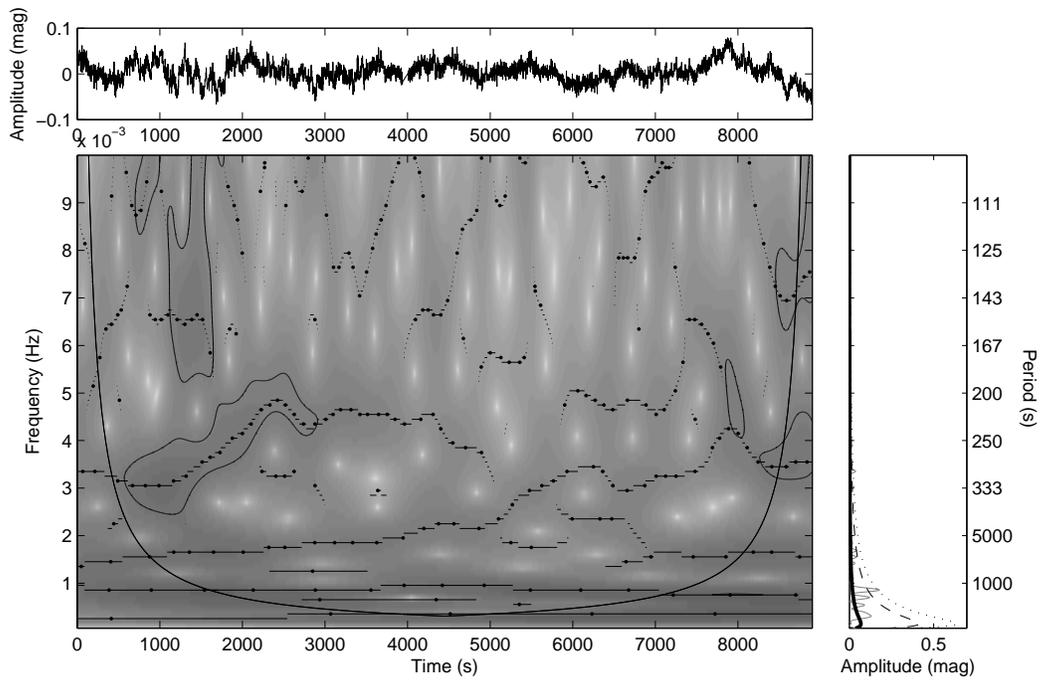}\\
  \caption{Prewhitened lightcurve, Morlet ridged scalogram and periodogram of run s0110. No oscillations are identified.}\label{fig.s0110_0_01.eps}
\end{figure}

\begin{figure}
  \includegraphics[width=6in]{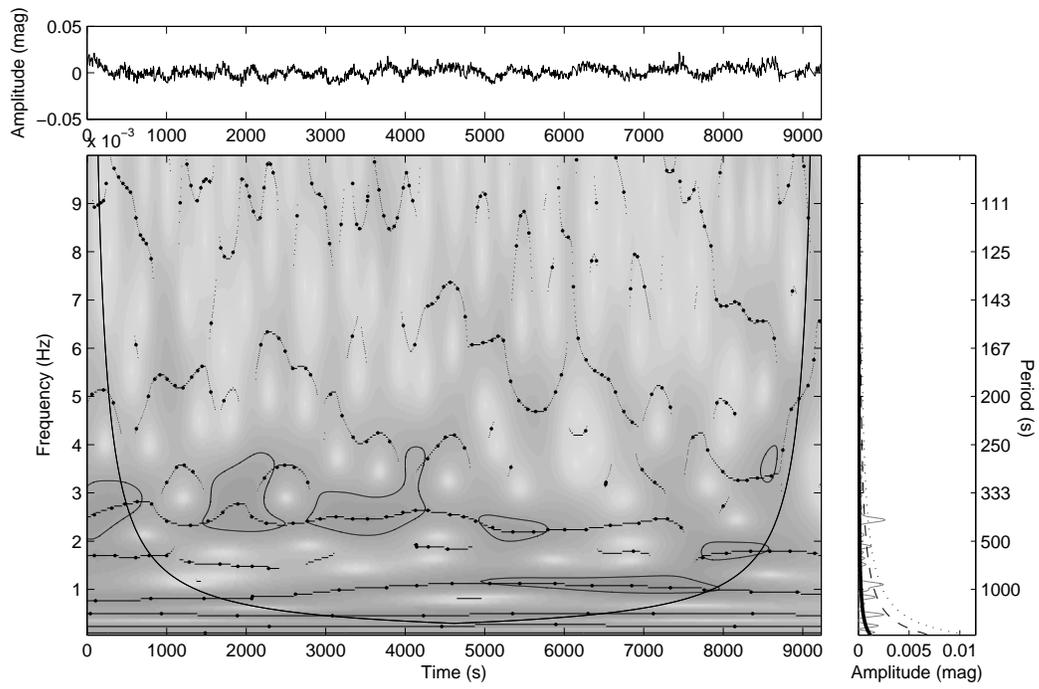}\\
  \caption{Prewhitened lightcurve, Morlet ridged scalogram and periodogram of run s0122, showing QPOs.}\label{fig.s0122_0_01.eps}
\end{figure}

Runs s0110 and s0122 are included to contrast the Morlet ridged scalograms of data with and without significant oscillations. Run s0110 (figure \ref{fig.s0110_0_01.eps}) shows no oscillations. Although there are regions of significant power, the instantaneous frequency fluctuates hugely, and hence these regions are not identified as indicating the presence of QPOs. In contrast, in s0122 (figure \ref{fig.s0122_0_01.eps}) we see in addition to the \un[410]{s} QPO identified in \citet{kn.Wo02a}  a \un[909]{s} oscillation persisting throughout most of the run,
especially prominent in the last half.

\begin{figure}
  \includegraphics[width=6in]{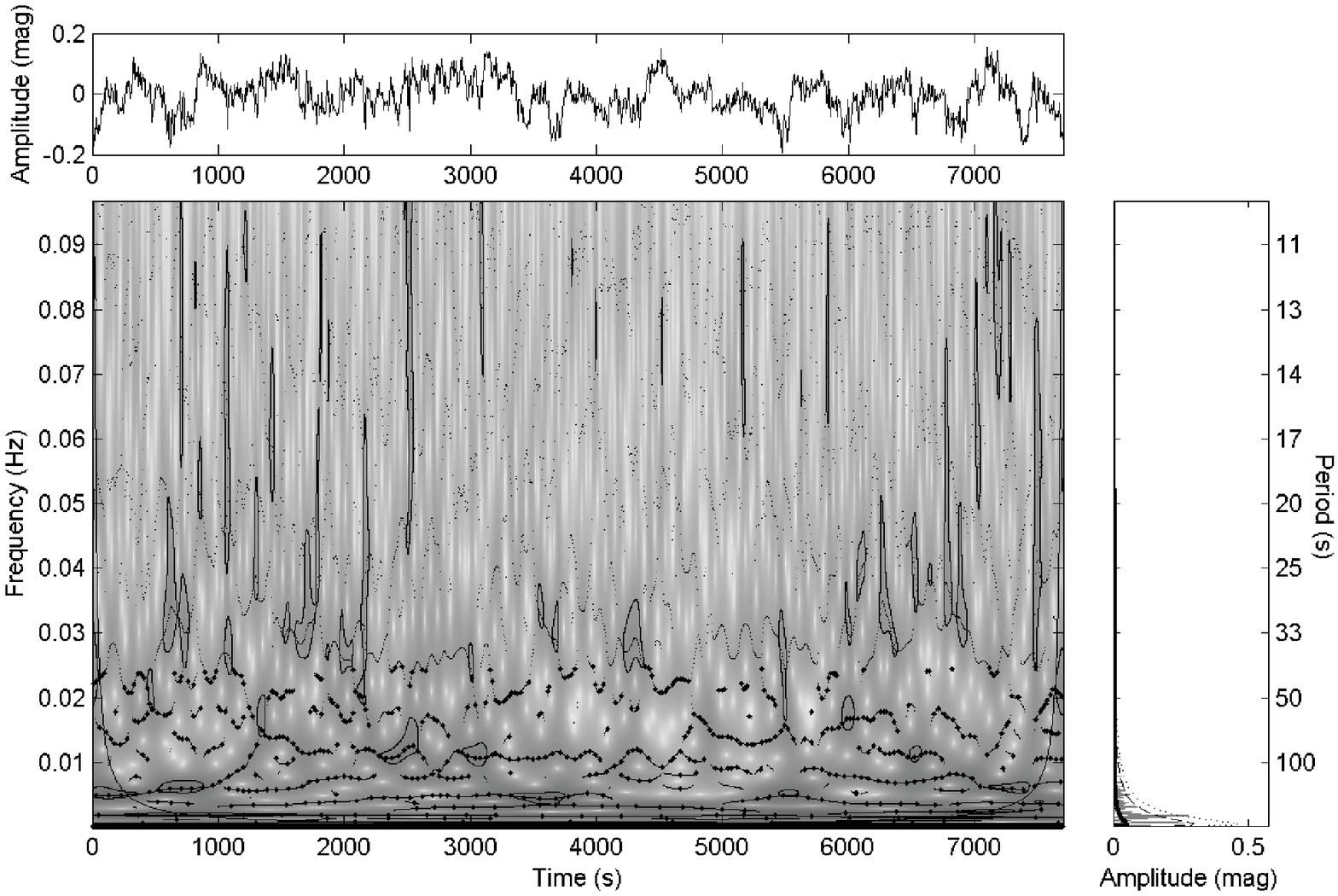}\\
  \caption{Prewhitened lightcurve, Morlet ridged scalogram and periodogram of run s0129.}\label{fig.s0129_0_1.eps}
\end{figure}

\subsubsection{s0129}

s0129 (figure \ref{fig.s0129_0_1.eps}) is a good example of the wavelet spectrum's use in indicating
oscillations of low coherence. In addition to the DNO (either DF or
lpDNO) at \un[93]{s} identified by \citet{kn.Wo02a}, there is
evidence of power at about \un[33]{s} throughout the scalogram
as well as three runs of QPOs with periods \un[578]{s}, \un[658]{s} and
\un[714]{s}. The most coherent run
lasts for only 5 cycles. Neither the \un[33]{s} nor the QPOs are visible in the power
spectrum.

\subsubsection{s1616}

\begin{figure}
  \includegraphics[width=6in]{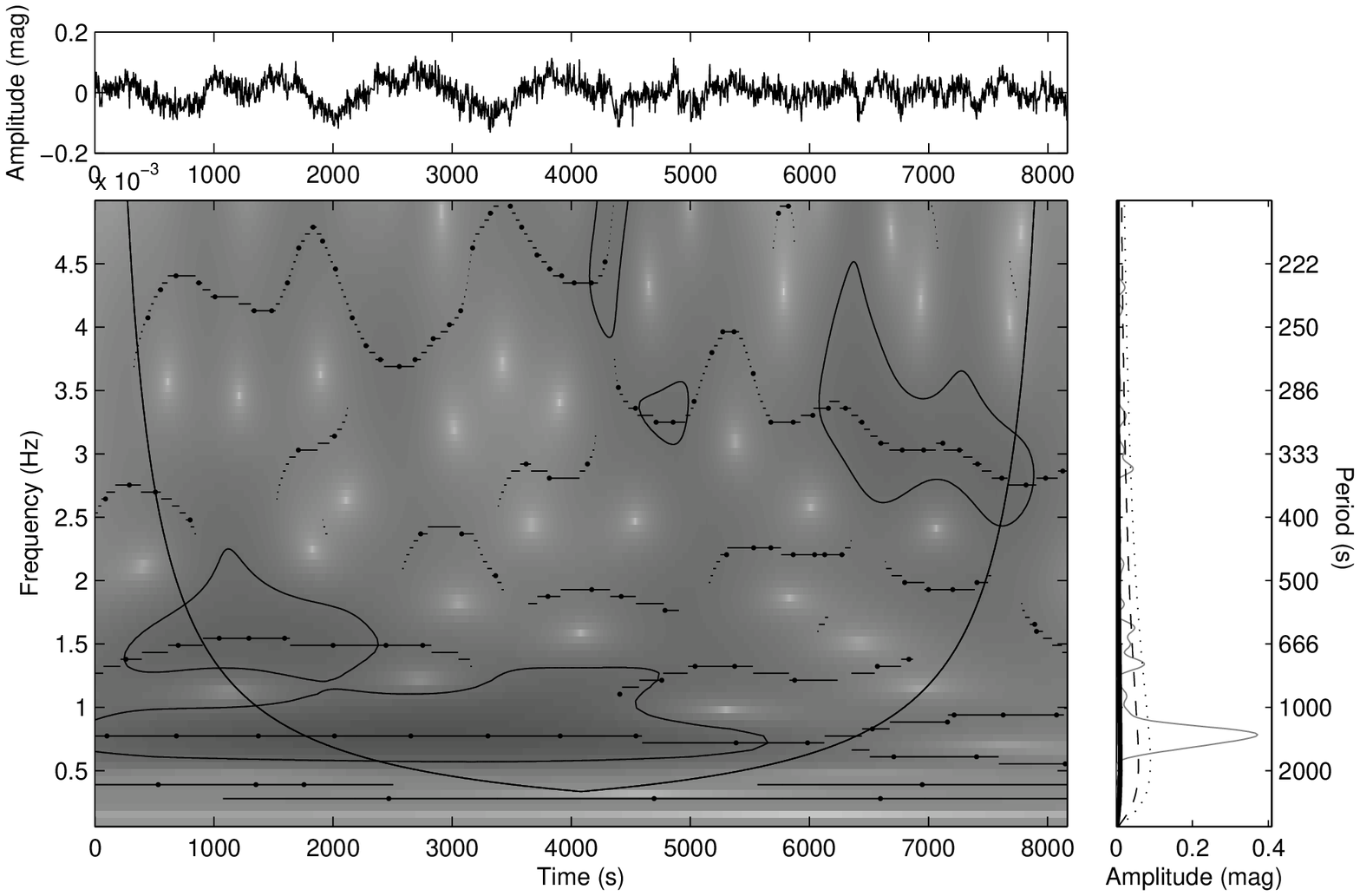}\\
  \caption{Prewhitened lightcurve, Morlet ridged scalogram and periodogram of run s1616.}\label{fig.s1616_0_005.eps}
\end{figure}

Run  s1616 was taken at the end of return to quiescence after a superoutburst, at
\un[0.912]{d}, and shows a typical triplet of QPO harmonics. A \un[694]{s} QPO appears twice in s1616, as well as a \un[1282]{s} QPO for the first half, and 7 cycles of
\un[333]{s} QPO, giving the triplet of QPO
fundamental, first and second harmonic. There is also evidence of an
\un[87]{s} oscillation near the end of the run (not shown in figure \ref{fig.s1616_0_005.eps}); this
may be either an lpDNO or the DNO fundamental.

\subsubsection{s2241}

\begin{figure}
  \includegraphics[width=6in]{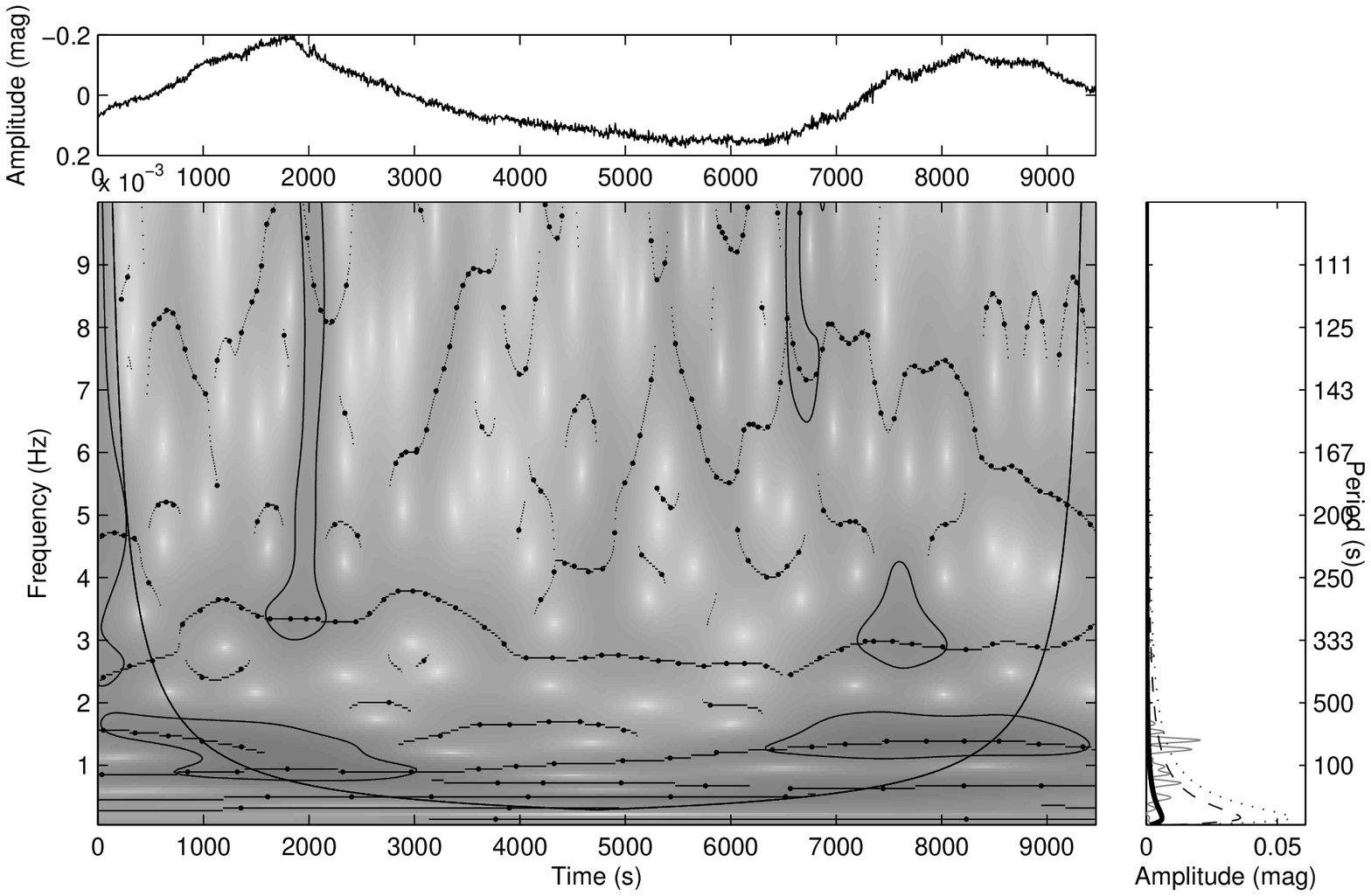}\\
  \caption{Raw lightcurve, Morlet ridged scalogram and periodogram of run s2241; note that the QPOs visible in the scalogram are only present during superhumps.}\label{fig.s2241_0_01.eps}
\end{figure}

Run s2241 (figure \ref{fig.s2241_0_01.eps}) was taken during superoutburst, and shows unclassified
QPOs with a period of about \un[700]{s}, first detected by \citet{kn.Wo02a}. The scalogram reveals the previously unrecognized fact that these QPOs are only present during the superhumps - in the top panel of figure \ref{fig.s2241_0_01.eps} we show the raw lightcurve of s2241, showing the superhumps, which can be matched with the QPOs in the scalogram below. The scalogram enables us to see clearly where the oscillations begin
and end, and for how many cycles they last. There is some evidence for a first harmonic at about \un[350]{s}, but this may be due to the non-sinusoidal profile of the QPOs.

Table \ref{Tbl:VWHyi_osc} summarizes the details of each oscillation
detected in our analysis; we include the run number, the type of
outburst, the start and end times of each oscillation (in seconds
from the beginning of the run), the period of the oscillation at
start and finish (in seconds), and the average amplitude (in
magnitudes). Oscillations showing
non-monotonic period changes were analysed in monotonic sections.
For oscillations with periods $>$\un[100]{s} the number of cycles
observed is included. DNOs are classified as fundamental (DF),
first(D1) or second harmonic(D2) or lpDNOs (DL). DNOs with periods
between 70 and \un[120]{s} may be either fundamental period DNOs, or
lpDNOs; we have decided to categorize all DNOs in this period range
as lpDNOs. DNO-related QPOs are classified as fundamental (QF),
first(Q1) or second harmonic (Q2) or subharmonic(QS). QPOs that do
not appear to be DNO-related are classified as QU, and QPOs from
quiescent light curves are classified as QQ. The second last column,
`Prev?', indicates whether the oscillation had been detected in
previous analyses (`y') or not (`n'), and the final column includes
comments.

\subsection{Results}

\begin{figure}
  \includegraphics[width=6in]{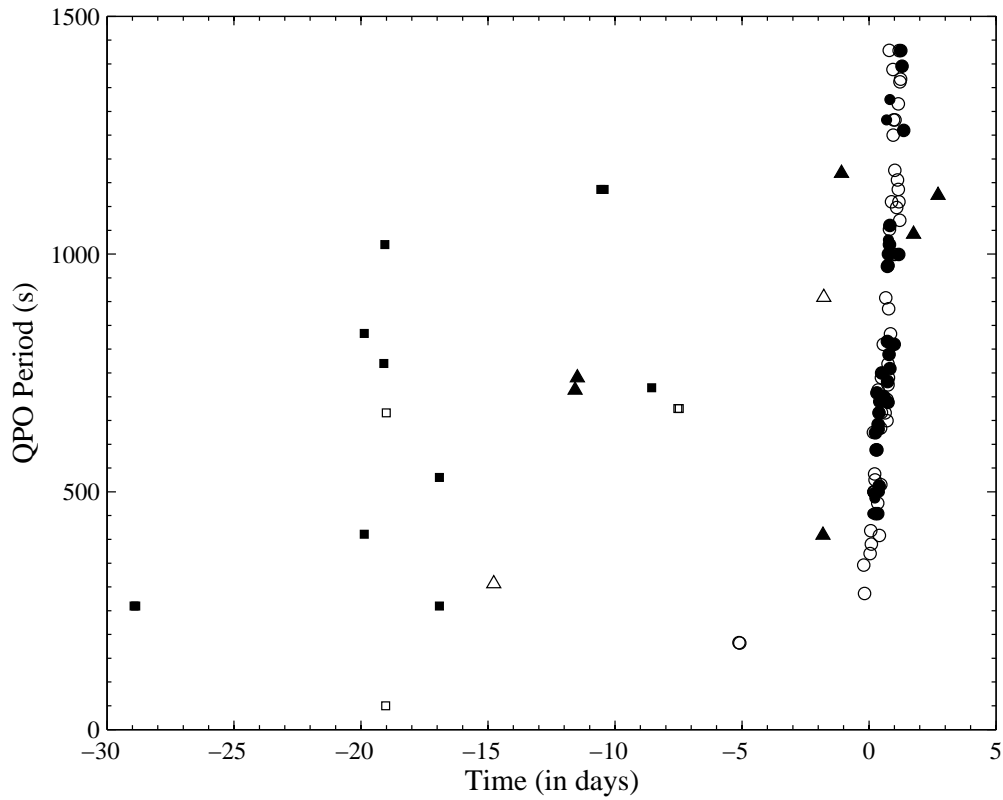}\\
  \caption{All QPOs analysed. Circles
indicate DNO-related QPOs, triangles indicate unclassified outburst
QPOs and squares indicate quiescent QPOs. Filled shapes are QPOs
previously analysed in Warner I, II or IV; open shapes are QPOs
added by this study.}\label{fig.all_qpos.eps}
\end{figure}

We have identified some 62 new QPOs to add the existing 44, showing that the Morlet ridged scalogram is a useful, objective tool for the detection of quasi-periodicities. Figure \ref{fig.all_qpos.eps} shows all the QPOs detected in our
analysis (for ease of viewing, DNO-related QPO harmonics are shown
at the implied fundamental). Previously studied QPOs
(\cite{kn.Wo02a}\ or \cite{kn.Wa06a}) are indicated by filled
shapes; open shapes indicate QPOs that have been detected using
wavelet analysis. Circles represent DNO-related QPOs, triangles
those QPOs which have no DNO counterpart and squares indicate
quiescent QPOs. Note that the large numbers of QPOs observed around $0<$T(d)$<1.15$ is at
least partly due to the observing campaign of Brian Warner and
Patrick Woudt, who selectively observed VW Hyi during late outburst.

All DNOs previously detected by \citet{kn.Wo02a} and \citet{kn.Wa06a} were detected in the Morlet scalogram, as well as 2 new DNOs. This is unsurprising: DNOs in general have higher coherence than QPOs, and are hence more likely to be detected in periodogram analysis such as that used in \citet{kn.Wo02a} and \citet{kn.Wa06a}. Out of 24 outburst runs showing DNOs, including all those covered in \citet{kn.Wa06a}, we found DNO-related QPOs in each run. In addition, after finding a triplet of QPOs in s1616, close inspection of the FT and wavelet showed evidence of DNOs, although of very low coherence.

In previous work, DNOs with periods $\geq$ \un[70]{s} have been
classified as DFs or lpDNOs almost arbitrarily; we have chosen to
classify every DNO with period $\geq$ \un[70]{s} as a lpDNO. Using
this classification, we find 8 previously observed lpDNOs, and we
have discovered a further 7.

\begin{deluxetable}{llllllllllp{2in}}
\tabletypesize{\small}
\tablewidth{8in}
\tablecaption{Table of QPOs detected in VW Hyi.\label{Tbl:VWHyi_osc}}
\rotate
\tablehead{\colhead{Name}  &  \colhead{Type}  &  \colhead{Osc}  &  \colhead{Osc end}  &  \colhead{Cycles}  &  \colhead{P start}  &  \colhead{P end}  &  \colhead{Mag}  & \colhead{Class}  &  \colhead{Prev?}  &  \colhead{Comments}\\ & & \colhead{start}&\colhead{end} & & & & & & &}
 \startdata
s0018  &  L  &  650  &  1900  &  4  &  346  &  346  &  0.008  &  QF & n &\\
s0018 & L & 1000 & 2000 &  & 75 & 75 & 0.007 & DL &  y  & \\
s0018 & L & 1500 & 2500 &  & 20 & 20 & 0.009 & DF &  y  & \\
s0018 & L & 3600 & 4600 & 3 & 286 & 286 & 0.007 & QF & n &\\
s0019 & L & 0 & 0 &  & 30 & 32 & 0.02 & D1 &  y & \\
s0019 & L & 1600 & 3200 & 3 & 487 & 487 & 0.05 & Q1 & y & \\
s0019 & L & 3500 & 6250 & 7 & 370 & 400 & 0.04 & Q1 & n & \\
s0019 & L & 5300 & 10000 &  & 20 & 20 & 0.02 & D2 &  y & \\
s0019 & L & 5300 & 10000 & 4 & 510 & 510 & 0.05 & Q1 &  y & \\
s0026 & L & 0 & 0 & 12 & 1042 & 1042 & 0.04 & QU &  y & \\
s0030 & L & 10000 & 0 & 3 & 1124 & 1124 & 0.04 & QU &  y & \\
s0111 & Super & 150 & 1300 & 4 & 307 & 307 & 0.007 & QU & n &\\
s0122 & Super & 0 & 5000 & 12 & 409 & 409 & 0.005 & QU &  y & \\
s0122 & Super & 2000 & 8000 & 7 & 909 & 909 & 0.003 & QU & n &\\
s0127 & Super & 0 & 2400 &  & 27.9 & 28.7 & 0.02 & DF &  y & \\
s0127 & Super & 1000 & 7000 & 5 & 1250 & 1250 & 0.03 & QS & n &\\
s0127 & Super & 2000 & 5000 & 7 & 454 & 454 & 0.02 & QF &  y & \\
s0127 & Super & 3600 & 4800 &  & 29.3 & 29.3 & 0.02 & DF &  y & \\
s0127 & Super & 5100 & 8200 &  & 29.7 & 30.5 & 0.03 & DF &  y & \\
s0127 & Super & 6000 & 8200 & 5 & 498 & 498 & 0.02 & QF &  y & \\
s0127 & Super & 8000 & 11000 & 3 & 1075 & 1075 & 0.03 & QS & n &\\
s0127 & Super & 9100 & 12000 &  & 32.2 & 32 & 0.015 & DF &  y & \\
s0127 & Super & 10000 & 12000 & 4 & 487 & 487 & 0.037 & QF &  y &\\
s0129 & S & 0 & 2500 &  & 33.5 & 33.5 & 0.02 & D2 & n & \\
s0129 & S & 0 & 2500 & 4 & 578 & 578 & 0.02 & Q1 & n & Implied first harmonic \\
s0129 & S & 2000 & 3500 &  & 93 & 93 & 0.03 & DL &  y & DF in Warner - may be DL \\
s0129 & S & 3000 & 5000 & 3 & 658 & 658 & 0.02 & Q1 & n & Implied first harmonic \\
s0129 & S & 3500 & 5000 &  & 93 & 93 & 0.019 & DL &  y & DF in Warner - may be DL \\
s0129 & S & 5000 & 8000 &  & 33 & 33 & 0.02 & D2 & n & \\
s0129 & S & 5000 & 8000 & 4 & 714 & 714 & 0.02 & Q1 & n & Implied first harmonic \\
s0484 & Super & 0 & 4000 &  & 30 & 32 & 0.02 & D1 &   y & \\
s0484 & Super & 0 & 3000 & 2.5 & 1282 & 1282 & 0.05 & QF & y & \\
s0484 & Super & 0 & 12000 & 5 & 2857 & 2757 & 0.05 & QS & n  & \\
s0484 & Super & 2000 & 6000 &  & 68 & 68 & 0.009 & DL & y & \\
s0484 & Super & 4000 & 6500 &  & 37 & 37 & 0.02 & D1 & y & \\
s0484 & Super & 4900 & 7500 & 6 & 488 & 488 & 0.03 & Q1 & y & \\
s0484 & Super & 5000 & 14000 & 7 & 1325 & 1325 & 0.05 & QF & y & \\
s0484 & Super & 6500 & 12000 &  & 34 & 34 & 0.02 & D1 & y & \\
s0484 & Super & 8000 & 14000 & 11 & 530 & 530 & 0.03 & Q1 & y & \\
s1307 & M & 0 & 1000 & 2 & 500 & 500 & 0.01 & QF & n  & \\
s1307 & M & 750 & 3000 &  & 31 & 31 & 0.005 & DF & n & \\
s1307 & M & 750 & 3000 & 5 & 250 & 250 & 0.005 & Q1 & n & \\
s1307 & M & 1500 & 3000 &  & 18 & 18 & 0.005 & D1 & y & \\
s1307 & M & 3500 & 4000 &  & 18 & 18 & 0.005 & D1 & y & \\
s1307 & M & 3500 & 4500 &  & 38 & 38 & 0.005 & DF & n & \\
s1307 & M & 3500 & 4500 & 5 & 175 & 175 & 0.01 & Q2 & n & \\
s1322 & M & 0 & 4000 &  & 31 & 31 & 0.01 & D2 & y  & \\
s1322 & M & 0 & 2500 & 11 & 357 & 357 & 0.05 & Q2 & n &  \\
s1322 & M & 3000 & 6000 & 4 & 476 & 476 & 0.05 & Q2 &   y  & \\
s1322 & M & 6000 & 10000 &  & 29 & 30 & 0.01 & D2 &   y & \\
s1322 & M & 6000 & 10000 & 8.5 & 465 & 465 & 0.05 & Q2 &  y & \\
s1322 & M & 10000 & 12000 &  & 32 & 32 & 0.01 & D2 &  y & \\
s1322 & M & 12000 & 16000 &  & 28 & 31 & 0.01 & D2 &  y & \\
s1322 & M & 12500 & 15200 & 6 & 420 & 420 & 0.02 & Q2 &  y & \\
s1571 & Super & 0 & 0 & 5.5 & 1170 & 1170 & 0.005 & QU &  y & \\
s1616 & Super & 0 & 2500 & 5 & 694 & 694 & 0.03 & Q1 & n & Implied fundamental\\
s1616 & Super & 0 & 5000 & 4 & 1282 & 1282 & 0.05 & QF & n & Implied fundamental\\
s1616 & Super & 6000 & 8000 &  & 87 & 87 & 0.01 & DL & n & \\
\tablebreak
s1616 & Super & 6000 & 8000 & 7 & 333 & 333 & 0.02 & Q2 & n & Implied fundamental - low coherence DNOs visible in periodogram\\
s2241 & Super & 0 & 2000 & 3 & 714 & 714 & 0.02 & QU &  y & \\
s2241 & Super & 5500 & 9400 & 5 & 740 & 740 & 0.02 & QU &  y & \\
s2243 & Super & 0 & 0 &  & 75 & 75 & 0.002 & DL &  y & \\
s2623 & M & 0 & 2000 &  & 22 & 22 & 0.015 & D2 &  y & \\
s2623 & M & 0 & 2000 & 5 & 408 & 408 & 0.02 & Q1 &  y & \\
s2623 & M & 1000 & 2500 & 6 & 244 & 244 & 0.03 & Q2 &  y & \\
s2623 & M & 2000 & 4000 &  & 25 & 25 & 0.014 & D2 &  y & \\
s2623 & M & 4000 & 6000 &  & 24 & 24 & 0.014 & D2 &  y & \\
s2623 & M & 4000 & 9000 & 3 & 1000 & 833 & 0.005 & QF &  y & Implied fundamental \\
s2623 & M & 5000 & 7000 & 7 & 263 & 263 & 0.036 & Q2 &  y & \\
s2623 & M & 7000 & 10000 & 13 & 253 & 253 & 0.035 & Q2 &  y & \\
\tablebreak
s2623 & M & 7000 & 10000 & 6 & 500 & 500 & 0.02 & Q1 & n  & \\
s2623 & M & 7500 & 10000 &  & 26 & 27 & 0.021 & D2 &  y & \\
s2623 & M & 10000 & 12500 &  & 25 & 27 & 0.019 & D2 &  y & \\
s2623 & M & 10000 & 12500 & 4 & 416 & 416 & 0.007 & Q1 & n & \\
s2915 & L & 700 & 1500 & 4 & 250 & 250 & 0.006 & Q1 & n  & \\
s2915 & L & 1500 & 2500 &  & 20 & 20 & 0.011 & D1 &  y & \\
s2915 & L & 1800 & 2300 & 2 & 250 & 250 & 0.006 & Q1 & n  & \\
s3078 & Super & 0 & 7000 &  & 14 & 14 & 0.004 & DF &  y & \\
s3078 & Super & 400 & 6000 &  & 90 & 90 & 0.002 & DL & n  & \\
s3078 & Super & 2000 & 4000 & 4 & 183 & 183 & 0.004 & QF & n & \\
s3078 & Super & 5500 & 6500 & 4 & 182 & 182 & 0.002 & QF & n & \\
s3416 & M & 0 & 1000 &  & 25 & 25 & 0.017 & D1 & y  & \\
s3416 & M & 0 & 1000 & 4 & 270 & 270 & 0.026 & Q2 & n & \\
s3416 & M & 1200 & 2500 &  & 27 & 27 & 0.015 & D1 & y  & \\
s3416 & M & 1200 & 2600 & 4 & 350 & 350 & 0.06 & Q1 & y  & \\
s3416 & M & 2600 & 4200 &  & 27 & 27 & 0.017 & D1 & y  & \\
s5248 & M & 0 & 4000 &  & 36 & 36 & 0.02 & D1 & y  & \\
s5248 & M & 0 & 0 &  & 96 & 96 & 0.02 & DL & n & \\
s5248 & M & 500 & 4000 & 6 & 454 & 454 & 0.08 & Q1 & n & \\
s5248 & M & 4000 & 8000 &  & 40 & 40 & 0.005 & D1 & y & \\
s5248 & M & 4000 & 8000 & 5 & 694 & 694 & 0.07 & QF & n & \\
s5248 & M & 5500 & 13000 & 5 & 1030 & 1030 & 0.05 & QF & y  & \\
s5248 & M & 5500 & 13000 & 3.5 & 2000 & 2000 & 0.06 & QS & y  & \\
s5248 & M & 8000 & 16000 &  & 40 & 40 & 0.005 & D1 & y  & \\
s6138-I & M & 0 & 3000 &  & 25 & 25 & 0.01 & D1 & y  & \\
s6138-I & M & 0 & 2200 &  & 25 & 26 & 0.01 & D1 & y  & \\
s6138-I & M & 0 & 2200 & 7 & 317 & 317 & 0.02 & Q1 & n & \\
s6138-I & M & 0 & 3000 & 3 & 1030 & 1030 & 0.03 & QS & n & \\
s6138-I & M & 3200 & 5000 & 3 & 667 & 667 & 0.05 & QF & n & \\
s6138-I & M & 3300 & 5000 &  & 26 & 28 & 0.01 & D1 & y & \\
s6138-I & M & 3300 & 5000 & 3 & 370 & 370 & 0.06 & Q1 & n & \\
s6138-I & M & 5000 & 7300 &  & 28 & 28 & 0.005 & D1 & y & \\
s6138-II & M & 0 & 3000 &  & 28 & 30 & 0.005 & D1 & y & \\
s6138-II & M & 0 & 5000 &  & 70 & 70 & 0.02 & DL & y & \\
s6138-II & M & 5000 & 14000 &  & 30.6 & 30.6 & 0.007 & D1 & y & \\
s6138-II & M & 5000 & 14000 & 13 & 344 & 392 & 0.04 & Q1 & y & \\
s6138-II & M & 6000 & 14000 & 5 & 1450 & 1450 & 0.02 & QS & n & \\
s6138-II & M & 7000 & 14000 & 11 & 769 & 769 & 0.04 & QF & n & \\
s6184 & L & 0 & 1000 &  & 24.8 & 27.8 & 0.02 & DF & y & \\
s6184 & L & 0 & 1800 & 5 & 370 & 370 & 0.01 & QF & n & \\
s6184 & L & 1000 & 2300 &  & 26 & 26 & 0.014 & DF & y & \\
s6184 & L & 1800 & 2300 &  & 83 & 83 & 0.01 & DL & y & \\
s6184 & L & 2300 & 3000 &  & 26 & 26 & 0.026 & DF & y & \\
s6184 & L & 2300 & 3800 & 3.5 & 418 & 418 & 0.024 & QF & n & \\
s6184 & L & 3000 & 3500 &  & 25 & 25 & 0.008 & DF & y & \\
s6184 & L & 3500 & 3800 &  & 26 & 26 & 0.014 & DF & y & \\
s6184 & L & 3800 & 5000 &  & 27 & 27 & 0.017 & DF & y & \\
s6184 & L & 3800 & 5000 &  & 74 & 74 & 0.016 & DL & y & \\
s6184 & L & 5000 & 5500 &  & 26 & 26 & 0.013 & DF & y & \\
s6184 & L & 5000 & 6000 & 2.5 & 390 & 390 & 0.01 & QF & n & \\
s6184 & L & 5500 & 6000 &  & 27 & 27 & 0.011 & DF & y & \\
s6316 & L & 0 & 2500 &  & 20 & 20 & 0.009 & D1 & y & \\
s6316 & L & 0 & 2000 & 6 & 294 & 294 & 0.001 & Q1 & y & \\
s6316 & L & 2500 & 4000 &  & 20 & 20 & 0.011 & D1 & y & \\
s6316 & L & 4000 & 6000 &  & 20.5 & 20.5 & 0.006 & D1 & y & \\
s6316 & L & 4000 & 6000 &  & 77 & 83 & 0.008 & DL & n & \\
s6316 & L & 4000 & 5000 & 5 & 227 & 227 & 0.015 & Q1 & y & \\
s6316 & L & 6000 & 10000 &  & 21.8 & 21.8 & 0.01 & D1 &  y & \\
s6316 & L & 10000 & 18000 &  & 21 & 23 & 0.016 & D1 &  y & \\
s6316 & L & 10000 & 15000 &  & 87 & 87 & 0.012 & DL & n &  \\
s6316 & L & 11000 & 18000 & 21 & 375 & 375 & 0.009 & Q1 &  y & \\
s6316 & L & 15000 & 18000 &  & 88 & 88 & 0.011 & DL & n &  \\
s6528-I & L & 0 & 6000 &  & 18 & 23 & 0.015 & D2 &  y & \\
s6528-I & L & 0 & 6000 & 13 & 666 & 666 & 0.024 & QF & n & Implied fundamental \\
s6528-I & L & 6000 & 12000 &  & 21 & 21 & 0.01 & D2 &  y & \\
s6528-I & L & 6000 & 12000 & 5 & 1299 & 1299 & 0.024 & QS & n & Implied fundamental\\
s6528-I & L & 12000 & 18000 &  & 23 & 23 & 0.011 & D2 &  y & \\
s6528-I & L & 12000 & 18000 & 6 & 885 & 885 & 0.013 & QF & n & Implied fundamental\\
s6528-I & L & 18000 & 20800 &  & 23 & 23 & 0.009 & D2 &  y & \\
s6528-I & L & 18000 & 20800 &  & 32 & 32 & 0.008 & D1 &  y & \\
s6528-I & L & 18000 & 20800 & 2 & 1052 & 1052 & 0.017 & QF & n & Implied fundamental\\
s6528-II & L & 0 & 6000 &  & 24 & 24 & 0.007 & D2 &  y & \\
s6528-II & L & 0 & 6000 & 21 & 370 & 370 & 0.015 & Q2 & n  & \\
s6528-II & L & 6000 & 10000 &  & 24 & 24 & 0.008 & D2 &  y & \\
s6528-II & L & 6000 & 10000 &  & 37 & 37 & 0.005 & D1 &  y & \\
s7222-I & Super & 0 & 2000 &  & 27 & 27 & 0.01 & D2 &  y & \\
s7222-I & Super & 0 & 10000 & 8 & 1282 & 1282 & 0.05 & QF & n & Implied fundamental\\
s7222-I & Super & 2000 & 4000 &  & 26 & 26 & 0.005 & D2 &  y & \\
s7222-I & Super & 2000 & 4000 & 5 & 625 & 476 & 0.02 & Q1 & n  & \\
s7222-I & Super & 3000 & 6000 &  & 27 & 29 & 0.011 & D2 &  y & \\
s7222-I & Super & 6000 & 7500 &  & 31 & 31 & 0.011 & D2 &  y & \\
s7222-I & Super & 6000 & 8000 &  & 38 & 38 & 0.013 & D1 &  y & \\
s7222-I & Super & 6000 & 7500 & 6 & 270 & 270 & 0.01 & Q2 &  y & \\
s7222-I & Super & 7000 & 9500 & 5 & 588 & 588 & 0.04 & Q1 & n  & \\
s7222-I & Super & 7500 & 9500 &  & 29 & 29 & 0.015 & D2 &  y & \\
s7222-II & Super & 0 & 2500 &  & 34 & 34 & 0.013 & D2 &  y & \\
s7222-II & Super & 0 & 2500 & 3 & 549 & 549 & 0.045 & Q1 & n  & \\
s7222-II & Super & 2500 & 4000 &  & 29 & 29 & 0.015 & D2 &  y & \\
s7222-II & Super & 4500 & 8000 & 3 & 1136 & 1136 & 0.038 & QF & n  & \\
s7222-II & Super & 5000 & 9000 & 8 & 333 & 333 & 0.035 & Q2 & y & \\
s7222-II & Super & 5000 & 10000 & 5 & 555 & 555 & 0.037 & Q1 & n & \\
s7222-II & Super & 5500 & 7500 &  & 28 & 28 & 0.022 & D2 & y & \\
s7222-II & Super & 7500 & 8500 &  & 30 & 30 & 0.016 & D2 & y & \\
s7222-II & Super & 8500 & 9000 &  & 32 & 32 & 0.009 & D2 & y & \\
s7222-II & Super & 9500 & 1000 &  & 34 & 34 & 0.01 & D2 & y & \\
s7301 & M & 0 & 2000 &  & 33 & 33 & 0.009 & D2 & y & \\
s7301 & M & 2000 & 4000 &  & 32 & 30 & 0.015 & D2 & y & \\
s7301 & M & 4000 & 6000 &  & 32 & 34 & 0.015 & D2 & y & \\
s7301 & M & 4000 & 7000 & 8 & 454 & 454 & 0.036 & Q2 & n & \\
s7301 & M & 6000 & 7000 &  & 29 & 29 & 0.014 & D2 & y & \\
s7301 & M & 7000 & 9000 &  & 32 & 32 & 0.012 & D2 & y & \\
s7301 & M & 7000 & 9000 & 5 & 456 & 456 & 0.05 & Q2 & n & \\
s7301 & M & 9000 & 0 &  & 29 & 31.8 & 0.014 & D2 & y & \\
s7311 & M & 0 & 0 &  & 90 & 90 & 0.014 & DL & n & \\
s7311 & M & 0 & 1000 & 3 & 312 & 312 & 0.014 & Q1 & y & \\
s7311 & M & 1800 & 3500 & 8 & 294 & 294 & 0.017 & Q1 &  y & \\
s7311 & M & 2000 & 3500 &  & 19 & 19 & 0.008 & D1 &  y & \\
s7311 & M & 3500 & 6000 &  & 20 & 20 & 0.006 & D1 &  y & \\
s7311 & M & 6000 & 15000 &  & 14 & 14 & 0.015 & D2 &  y & \\
s7311 & M & 6000 & 7400 &  & 20 & 20 & 0.007 & D1 &  y & \\
s7311 & M & 6000 & 10000 & 12 & 316 & 316 & 0.008 & Q1 &  y & \\
s7311 & M & 7000 & 9000 & 4 & 476 & 476 & 0.028 & QF & n & Implied fundamental\\
s7311 & M & 7400 & 10000 &  & 20 & 21 & 0.006 & D1 &  y & \\
s7311 & M & 10000 & 13000 & 9 & 333 & 333 & 0.028 & Q1 &  y & \\
s7311 & M & 10500 & 12500 &  & 21 & 21 & 0.01 & D1 &  y & \\
s7311 & M & 12500 & 0 &  & 21 & 22 & 0.009 & D1 &  y & \\
s7311 & M & 13000 & 16000 & 6 & 345 & 345 & 0.015 & Q1 &  y & \\
s7311 & M & 13000 & 16000 & 4 & 666 & 666 & 0.016 & QF &  y & \\
s7342 & M & 0 & 0 &  & 20 & 20 & 0.006 & D1 &  y & \\
s7342 & M & 0 & 2000 & 7 & 236 & 236 & 0.009 & Q2 &  y & \\
s7342 & M & 1800 & 6000 & 5 & 714 & 714 & 0.006 & QF & n  & \\
s7342 & M & 2000 & 6000 & 10 & 321 & 321 & 0.01 & Q1 & y & \\
s7342 & M & 6000 & 0 &  & 14 & 14 & 0.015 & D2 & y & \\
s7342 & M & 6000 & 8000 &  & 38 & 38 & 0.009 & DF & y & \\
s7342 & M & 6000 & 10000 & 14 & 136 & 136 & 0.005 & Q2 & n & \\
s7342 & M & 8000 & 9200 & 5 & 256 & 256 & 0.015 & Q1 & y & \\
s7342 & M & 8000 & 10000 & 4 & 500 & 500 & 0.009 & QF & y & \\
s7621 & L & 0 & 600 &  & 28 & 28 & 0.007 & DF & y & \\
s7621 & L & 600 & 2000 &  & 27 & 30 & 0.011 & DF & y & \\
s7621 & L & 4000 & 6000 & 5 & 454 & 454 & 0.012 & QF & y & \\
s7621 & L & 6000 & 11000 & 11 & 454 & 454 & 0.027 & QF & n & \\
s7621 & L & 8000 & 10000 &  & 31 & 33 & 0.008 & DF & y & \\
s7621 & L & 11000 & 14000 &  & 18 & 18 & 0.01 & D1 & y & \\
s7621 & L & 11000 & 14000 &  & 38 & 38 & 0.013 & DF & y & \\

\enddata
\end{deluxetable}

\section{Discussion and Conclusion}
\label{s.conclusion}

Lightcurves that contain multiple non-stationary components cannot
be satisfactorily analysed using the periodogram alone, as the
periodogram obscures the time-varying nature of the signal.
Time-frequency representations allow signals to be viewed in the
time and frequency domains simultaneously, enabling non-stationary
components to be analysed. Various TFRs have been used in astronomy,
but only for signals that have very high signal-to-noise ratios,
cover a narrow frequency band, and have only a few components. For
signals such as those found in cataclysmic variable stars, which
contain multiple quasi-periodicities over a broad frequency range,
against high background noise, it is not clear which of the TFRs
used in the literature (if any) is most appropriate.

We have investigated the time-frequency resolution of six TFRs,
using three synthetic signals mimicking the key features of our
data: broad frequency range, multiple components (some close
together in frequency), amplitude modulation and intermittency. Interference terms are
produced in many TFRs, so for our data we had to tune the TFRs to
completely attenuate ITs, which could otherwise be mistakenly
identified as signal terms. We found that the CWD could not be tuned
to completely remove ITs for signals with components that overlap in
time, and hence was not appropriate for our data. The Gabor spectrogram cannot provide simultaneously high time and frequency resolution over the wide frequency range of our data. While both the CKR
and the ASPWVD gave good time and frequency resolution over a range
of frequencies, they could not completely attenuate ITs at very high
frequencies; the only TFR with achieved this was the wavelet
scalogram. However, at high frequencies the wavelet scalogram, when
tuned for high time resolution, did not have good frequency
resolution. By computing the wavelet ridges of the Morlet scalogram,
however, this shortcoming was addressed. We used the maxim points of the Mexican Hat scalogram to enable us to detect the extrema of our data.

We also reviewed the statistical properties of the six TFRs in
the presence of noise, and found that the wavelet scalogram was the
minimum variance estimator for affine TFRs in the presence of
non-white noise, with a statistical distribution easily computable
from the power spectrum.

We analysed archival data of VW Hyi with the Morlet ridged scalogram, and compared our results with those previously published. We found that the Morlet ridged scalogram could be used to reliably detect
oscillations of only a few cycles ($<$5) (eg s1307, s6138 and
s6184) and we used this feature to identify QPOs
objectively against criteria based on the appearance of QPOs
identified in previous analyses. We added some 62 QPOs to the existing 44 of VW
Hyi using this method.

For low amplitude coherent oscillations at short periods (ie $<$
\un[50]{s}), it is no surprise that the periodogram is still the
best means of detection. However, for oscillations in this range that have phase changes
or gaps (e.g. s0129, s1307, s6528), the Morlet ridged scalogram can be used
to determine the structure of the oscillation, but only for
amplitudes $>$ \un[0.005]{mag}. The Morlet ridged scalogram can also be used
to detect short period oscillations of low coherence, which are not
visible in the periodogram (e.g. s0129).

Oscillations with periods greater than \un[800]{s} (eg
s0127, s0484, s1322, s6138) fall in a region of the periodogram that
often shows considerable noise in CV data, making differentiation
between signal and noise difficult, especially if the amplitude of
the oscillation is small, and they have low coherence. However, the statistical properties of the scalogram enabled us to fit a model to the noise continuum and hence detect significant periodicities.

The Morlet ridged scalogram also provided a clear picture of the time evolution of different frequency components, providing insights unavailable from the raw lightcurve or periodogram. For example, groups of spikes in
the periodogram were resolved as either a single oscillation with
discrete period changes (e.g. s0019), or as multiple simultaneous oscillations of similar period
(e.g. s2915, s2623, s3416 and s1322). While
the QPOs in s2241 had been observed previously, the wavelet scalogram
showed that they only occur during superhumps, which was not
previously recognised. We also observed further links between DNO and QPO
behaviour, such as period tracking and simultaneous
initiation/termination of oscillations (e.g. s2623).

\subsection{Software}

The figures of all TFRs in this paper except wavelet scalograms were generated using the Matlab
Time-Frequency Toolbox, developed by Fran\c{c}ois Auger, Olivier
Lemoine, Paulo Gon\c{c}alv\`{e}s and Patrick Flandrin. It is
available at http://tftb.nongnu.org/, and is also compatible with
GNU Octave (available at http://www.gnu.org/software/octave/). An
Ansi C version is available at
http://www-lagis.univ-lille1.fr/$\sim$davy/\-toolbox\-/Ctftbeng.html.

For the wavelet scalograms, we used the Matlab toolbox of \citet{kn.To98} as a starting
point (available at http://paos.colorado.edu/research/wavelets/software.html), but have written additional instantaneous frequency algorithms and confidence contour functions. All software is available on request.

\subsection{Acknowledgements}
The author acknowledges support from the University of Cape Town and from the National  Research Foundation of South Africa. I thank Brian Warner, Chris Koen, Patrick Woudt and Michael Mandler for valuable discussions and Laszlo Kiss for helpful comments on a first draft of this work.


\end{document}